\documentclass[reprint,eqsecnum,floats,aps,amsmath,amssymb,nofootinbib,prd,onecolumn, showpacs]{revtex4-1}

\usepackage{graphicx,physics}
\usepackage{amsmath,amssymb,mathtools,mathrsfs}
\usepackage{hyperref}
\usepackage{graphicx}
\usepackage{subfigure}
\usepackage{arydshln}
\usepackage{xcolor}
\usepackage{braket}
\usepackage{tensor}
\usepackage{enumitem,array,textcomp}

\usepackage{color}

\begin{document}
	
	\title{The time-dependent mass of cosmological perturbations in loop quantum cosmology: Dapor--Liegener regularization}
	\affiliation{Instituto de Estructura de la Materia, IEM-CSIC, Serrano 121, 28006 Madrid, Spain}
	\author{Alejandro Garc\'ia-Quismondo}
	\email{alejandro.garcia@iem.cfmac.csic.es}
	\affiliation{Instituto de Estructura de la Materia, IEM-CSIC, Serrano 121, 28006 Madrid, Spain}
	\author{Guillermo  A. Mena Marug\'an}
	\email{mena@iem.cfmac.csic.es}
	\author{Gabriel S\'anchez P\'erez}
	\email{gabriel.sanchez@iem.cfmac.csic.es}
	\affiliation{Instituto de Estructura de la Materia, IEM-CSIC, Serrano 121, 28006 Madrid, Spain}
	\begin{abstract}
In this work, we compute the time-dependent masses that govern the dynamics of scalar and tensor perturbations propagating on an effective flat, homogeneous, and isotropic background within the framework of loop quantum cosmology, regularized according to the procedure put forward by Dapor and Liegener. To do so, we follow the two main approaches that, in the field of loop quantum cosmology, lead to hyperbolic equations for the perturbations in the ultraviolet sector: the hybrid and dressed metric formalisms. This allows us to compare the masses resulting from both proposals and analyze their positivity in regimes of physical interest: the big bounce and the contracting de Sitter phase in the asymptotic past that is a defining feature of the model under consideration.
	\end{abstract}
	
	\pacs{04.60.Pp, 04.60.Kz, 98.80.Qc.}
	\maketitle					

\section{Introduction}\label{sec:Intro}

General relativity \cite{Wald,Hawking-Ellis} and quantum mechanics \cite{CT,Galindo} are the two cornerstones of modern theoretical physics and have made possible the comprehension of an extremely wide spectrum of physical phenomena in the past century. Their partial merger in the form of quantum field theory in curved spacetime \cite{Birrel-Davies,WaldQFTCS} has served, among others, to formulate a theory of cosmological perturbations \cite{perturbations1,perturbations2,HalliwellHawking,ShiraiWada} that has proven to be largely successful in the description of the primordial seeds of the large scale structure (LSS) in the present Universe. Indeed, the study of perturbations around a Friedmann-Lema\^itre-Robertson-Walker (FLRW) cosmology, together with the inflationary paradigm \cite{inflationmodels1,inflationmodels2,inflationmodels3,inflationmodels4,inflation1,inflation2}, have led to a surprisingly accurate description of the observations, for instance regarding the cosmic microwave background (CMB) \cite{data1,data2,data3}. However, even though the predictions of the standard cosmological model have been mostly successful, there might exist certain tensions in the region of moderately large scales \cite{data2}. Although these tensions may be a statistical realization of cosmic variance, there is a belief that the accumulated significance of anomalies in the confrontation between theory and observation might be due to the fact that, in the standard cosmological model, the spacetime geometry on which the perturbations propagate is treated purely at the classical level, as described by general relativity \cite{ashrec,agullrec}. Nevertheless, this classical theory is known to be fundamentally incomplete, in the sense that it predicts its own breakdown in the form of singularities \cite{Hawking-Ellis}. For this reason, there is an interest in constructing a theory of cosmological perturbations where the quantum nature of gravity is appropriately taken into account and {in} analyzing how this affects the predictions for the CMB and the LSS, that provide a promising probe of the physics of the very early Universe and, thus, a testbed for quantum geometry effects.

Among the candidates for a quantum theory of gravity, loop quantum gravity (LQG) stands out as a solid approach that has undergone a major development in the past decades \cite{Thiem,ALQG}. It is a background independent, nonperturbative canonical quantization of general relativity in 3+1 spacetime dimensions, based on an adaption of strategies employed in Yang-Mills theories to the gravitational degrees of freedom. Although the quantization program remains incomplete as of yet, it has successfully been applied to cosmological scenarios, where the high number of symmetries makes it possible to overcome some of the difficulties that arise in the full theory. This led to the birth of the field of loop quantum cosmology (LQC) \cite{LQC,AS}. Although the formalism of LQC is affected by some mathematical ambiguities, a number of results have been found to be robust. The most outstanding one is the resolution of the big bang singularity. In FLRW spacetimes \cite{MMO,APS1,APS2}, the standard cosmological singularity is replaced with a quantum bounce, that joins deterministically two classical branches, one in contraction and the other in expansion. Despite the success of the homogeneous and isotropic description, it is known that such a model is insufficient to describe the Universe we inhabit. Indeed, inhomogeneities are to be considered in order for any structure to be formed by gravitational instability. This naturally drove the community to the introduction of inhomogeneities in the FLRW models in the form of perturbations. In this regard, mainly two different paths have been developed in extent leading to predictions compatible with observations: the so-called hybrid \cite{inflationaryuniverse,inflationarymodel,MSvariables,GIper,quantumcorrectionsMSeq,Olm} and dressed metric \cite{dressed1,dressed2,dressed3,dressed4} formalisms.

Both approaches to inhomogeneous LQC are based on the assumption that there exists a physical regime between the fully quantum one and the domain of validity of semiclassical descriptions where the quantum gravity effects mainly affect the homogeneous sector of the cosmological system \cite{inflationaryuniverse}. With this motivation in mind, both approaches seek to implement a quantization program based on the choice of two different representations: one of a quantum gravity nature, to describe the homogeneous sector of the model, and a more standard one, fit to describe the perturbative inhomogeneities. Nonetheless, both formalisms differ in the formal construction of the proposal itself. On the one hand, the hybrid approach regards the entire cosmological system as a constrained symplectic manifold, obtained from the truncation of the action at the lowest nontrivial order in the perturbations (around FLRW spacetimes with compact sections) \cite{GIper}. As such, the background homogeneous cosmology and the perturbations are treated on a similar footing.  On the other hand, the dressed metric approach deals with the quantization of the system in two steps: it treats the homogeneous sector first, obtaining the corresponding dynamical trajectories of the background in a kind of mean field approximation that dresses the FLRW metric with quantum effects, and then lifts these trajectories to the truncated phase space that incorporates the inhomogeneities. From the theoretical point of view, these two procedures are in contrast with each other. Hence, even though the two approaches share a number of common features, such as the ultraviolet behavior of the perturbations \cite{inflationaryuniverse,inflationarymodel,dressed2}, differences are to be expected. For instance, such differences arise in the time-dependent masses appearing in the dynamical equations of scalar and tensor perturbations. A comparison between the masses resulting from the hybrid and dressed metric approaches in the context of effective LQC has already been carried out in Ref. \cite{positividad}. In particular, the corresponding time-dependent masses at the instant of the bounce were discussed in that work. It was shown that, whereas the hybrid masses turn out to be positive for a certain class of potentials in the scenarios of kinetic dominance at the bounce (these scenarios are the most interesting ones from the point of view that they allow a good fit of the observed CMB spectra while retaining quantum effects at large scales), the dressed metric masses are negative in those cases. These differences regarding the effective masses at the bounce are quite important. Indeed, the instant of the bounce, at which the physical volume of the Universe attains its minimum, is often viewed as providing a preferred choice of time to set initial conditions for the perturbations (typically understood as defining an initial vacuum state). Whether or not these initial conditions are well defined often depends on the properties of the effective masses at the initial time and, more specifically, on their positivity. This is the case, for instance, if one wishes to define adiabatic states for all wavelengths \cite{dressed2,dressed3,dressed4,CMBhybrid,CMBdressed,DaniJavi}.

Recently, an increasing attention has been devoted to one of the mathematical ambiguities of the formalism of LQC: the regularization and subsequent definition of the Hamiltonian constraint \cite{ma}. Dapor and Liegener have put forward a regularization procedure that follows the ideas of LQG more faithfully than the standard procedure that had been used in the community since its inception \cite{DL0,DL,DLdetailed}. Traditionally, the Hamiltonian constraint has been defined in LQC by exploiting the symmetries of homogeneity and spatial flatness. When these properties are exhibited by the system under consideration, the two parts that compose the Hamiltonian in general relativity (namely, the Euclidean and Lorentzian parts, the former containing all the contributions if the spacetime signature were positive) turn out to be proportional to each other. Therefore, a symmetry reduction can be performed on the Hamiltonian before the regularization procedure is implemented, in such a way that it can be written in terms of the Euclidean contribution alone. As a result, regularizing this contribution suffices to define the entire Hamiltonian in a way such that it can be readily quantized. This is conceptually very different from the procedure followed in the full theory, where the absence of these symmetries requires that the Euclidean and Lorentzian parts be treated individually. It is in this sense that the Dapor--Liegener proposal resembles LQG more closely: it regularizes the Euclidean and Lorentzian contributions separately without relying on the symmetries of the system under consideration. Not only does {it} seem theoretically more satisfactory, but also it may help understanding whether the results obtained within the standard approach to LQC are robust. This objective has inspired a number of studies (see, for instance, Refs. \cite{Paramc1,Paramc2,genericness,Agullo,Haro,MMODL,DLBI,DLhLQC}), the aim of which was to explore the features of the model, how they compare to the standard ones, and the implementability of the procedure in more complicated scenarios, among others. It has been found that, although the big bang singularity is still resolved by a bouncing mechanism, the resulting bounce is quantitatively and qualitatively different from the standard one. Indeed, instead of joining two classical universes in a symmetric fashion, the new big bounce is asymmetric, inasmuch as it joins an FLRW branch with an asymptotically de Sitter one, where an emergent cosmological constant of Planck order appears. With this caveat, the big bang resolution in homogeneous and isotropic LQC is found to be robust and not an artifact of a specific regularization of the scalar constraint.

The Dapor--Liegener regularization scheme has been implemented in isotropic and anisotropic scenarios \cite{DLBI}. The introduction of perturbations within the dressed metric approach has already been considered in Ref. \cite{Agullo}. In a recent work \cite{DLhLQC}, the particularization of the formalism of hybrid LQC to the Dapor--Liegener regularization of the FLRW cosmological background has been studied in detail, and two different admissible prescriptions have been put forward for the quantum definition of certain geometrical operators. At this point, it seems natural to wonder how the predictions of the hybrid and dressed metric formalisms compare to each other when the homogeneous geometry is regularized following the proposal of Ref. \cite{DL}. Furthermore, a new regime of interest arises in this setting: not only is it interesting to study the properties of the time-dependent masses at the instant of the bounce, but also in the new de Sitter epoch that emerges with this regularization scheme. Indeed, the asymptotic past seems a good candidate to try and set initial conditions for the perturbations, given that a de Sitter regime is very rapidly approached before the bounce and the Bunch--Davies vacuum may be a natural choice of vacuum in this scenario \cite{Agullo}. This paper aims to address these questions, by computing and analyzing the time-dependent masses seen by the perturbations in the hybrid and dressed metric formalisms when the homogeneous geometry is regularized according to the Dapor--Liegener proposal. Concretely, we will study and compare the masses in full detail in the two commented regions of physical interest to set initial conditions: the big bounce and the asymptotic de Sitter regime.

This paper is structured as follows. In Sec. \ref{sec:hybrid}, we overview the most important elements of effective hybrid LQC and provide the expression of the time-dependent masses that govern the propagation of scalar and tensor perturbations, distinguishing between the two possible prescriptions discussed in Ref. \cite{DLhLQC} for the definition of the geometric operators that are necessary to determine these masses. Then, in Sec. \ref{sec:dressed}, we obtain the corresponding expression of those masses in the dressed metric approach, so as to be in a position to compare the results of both formalisms. As commented above, this comparison is especially enlightening in what regards the positivity in some interesting physical regimes. In Sec. \ref{sec:bounce}, the considered masses are evaluated at the instant of the bounce and their positivity is studied. A similar analysis is carried out in the asymptotic de Sitter regime in Sec. \ref{sec:dS}. Finally, we summarize and discuss the main results in Sec. \ref{sec:conclusion}. Throughout this work, we set the speed of light and the reduced Planck constant equal to one.

\section{The hybrid approach}\label{sec:hybrid}

Let us briefly review the aspects of the hybrid approach to inhomogeneous LQC that are relevant for the derivation of the time-dependent masses of the gauge invariant perturbations. We will focus our attention on quantum states of the FLRW cosmological background for which the effective description of homogeneous and isotropic LQC is valid. This effective description can be attained in practice by replacing the dependence on the Hubble parameter of the relevant geometric operators with sinusoidal functions of that parameter, as a consequence of the regularization of the quantum expressions by means of holonomy elements \cite{AS,taveras,ACA}.  Therefore, we omit any details about the fully quantum formalism for the sake of brevity and refer directly to the effective counterpart of all quantities. For a more thorough discussion, we refer the reader to Refs. \cite{inflationaryuniverse,inflationarymodel,MSvariables,GIper,quantumcorrectionsMSeq,Olm,DLhLQC}.  

The hybrid formalism for cosmological perturbations is based on a gauge invariant description of an inhomogeneous system, regarded as a perturbation around a (spatially compact) homogeneous and isotropic cosmology, and derived from the truncation of the Hilbert-Einstein action at quadratic order in the perturbations. The resulting truncated system is subject to a number of constraints, that ensure the covariance at the considered order of truncation \cite{GIper}. This constrained system can be cast in a canonical form and then be quantized following the ideas of Dirac \cite{Dirac} by adopting an LQC representation for the homogeneous sector and a standard Fock quantization for the gauge invariant perturbations. In the remainder of this paper, we deal with perturbations around a flat FLRW cosmology, with spatial sections that have the topology of a three-torus. We consider a matter content given by a minimally coupled scalar field subject to a certain field potential, so as to induce nontrivial cosmological dynamics. 

The canonical variables employed to coordinatize the phase space of our perturbed system are the following. The homogeneous sector that describes the FLRW background, that is treated exactly in the perturbation hierarchy, can be described by a variable $\tilde{\alpha}$ corresponding to the logarithmic scale factor and by its canonical momentum $\pi_{\tilde{\alpha}}$, together with another canonical pair, $\tilde{\varphi}$ and $\pi_{\tilde{\varphi}}$, associated with the zero mode of the scalar field and its momentum \cite{GIper,DLhLQC}. As far as the inhomogeneous sector is concerned, in absence of vector matter fields, vector perturbations are pure gauge and thus we will ignore them. As regards the scalar perturbations, the relevant gauge invariant variables are the so-called Mukhanov--Sasaki modes \cite{sasaki,kodamasasaki,mukhanov}, $v_{\vec{n},\epsilon}$ and $\pi_{v_{\vec{n},\epsilon}}$, where $\vec{n}\in \mathbb{Z}^3-\{0\}$  is the wavevector, with its first nonvanishing component being positive, and $\epsilon=\pm$ is the parity (the passage to a continuum of Fourier modes is attained in a suitable limit, as explained in Ref. \cite{beamm}). The rest of the degrees of freedom in the scalar perturbations are gauge and can be assigned to perturbative constraints, conveniently Abelianized at our order of perturbative truncation, or to variables canonically conjugated to those constraints \cite{GIper}. The tensor perturbations, on the other hand, can be described by  a series of mode coefficients $\tilde{d}_{\vec{n},\epsilon,\tilde{\epsilon}}$ and $\pi_{\tilde{d}_{\vec{n},\epsilon,\tilde{\epsilon}}}$, analogous to the Mukhanov--Sasaki ones (they are also gauge invariant, in the sense of the Bardeen potentials \cite{bardeen}),  but with an additional label $\tilde{\epsilon}=+,\times$ that refers to the polarization of the mode. These variables satisfy a global Hamiltonian constraint that is composed by a homogeneous contribution and a number of terms which are quadratic in the gauge invariant perturbations. This global Hamiltonian can be written as
\begin{align}
H=\dfrac{e^{-3\tilde{\alpha}}}{2}\left(2e^{3\tilde{\alpha}}H_{|0}-\Theta_o^{S}\pi_{\tilde{\varphi}}-\Theta_e^S-\Theta^T\right),\label{H}
\end{align}
where
\begin{align}
	H_{|0}&=\dfrac{e^{-3\tilde{\alpha}}}{2}\left(\pi_{\tilde{\varphi}}^2-\mathcal{H}_0^{(2)}\right),\quad
	\mathcal{H}_0^{(2)}=\pi_{\tilde{\alpha}}^2-2e^{6\tilde{\alpha}}\bar{W}, \label{H02}\\
	\Theta_o^S&=-\vartheta_o\sum_{\vec{n},\epsilon}(v_{\vec{n},\epsilon})^2,\\
	\Theta_e^S&=-\sum_{\vec{n},\epsilon}[(\vartheta_e \omega_n^2+\vartheta_e^q)(v_{\vec{n},\epsilon})^2+\vartheta_e(\pi_{v_{\vec{n},\epsilon}})^2],\\
	\Theta^T&=-\sum_{\vec{n},\epsilon,\tilde{\epsilon}}[(\vartheta_e \omega_n^2+\vartheta_T^q)(\tilde{d}_{\vec{n},\epsilon,\tilde{\epsilon}})^2+\vartheta_e(\pi_{\tilde{d}_{\vec{n},\epsilon,\tilde{\epsilon}}})^2].
\end{align}
Here, $\omega_n^2=-4\pi^2|\vec{n}|^2/l_0^2$, $l_0$ is the coordinate length of the fundamental cycles of the three-toroidal sections (that can be fixed freely), and $\bar{W}$ is related to the scalar field potential $W$ via $\bar{W}(\tilde{\varphi})=\sigma^4W(\tilde{\varphi}/\sigma)$, where $\sigma^2=4\pi G/3l_0^3$ and $G$ is the Newtonian gravitational constant. Finally, the $\vartheta$-functions are 
\begin{align}	\label{varthetae}
	\vartheta_e&=e^{2\tilde{\alpha}}, \quad \vartheta_o=-12e^{4\tilde{\alpha}}\bar{W}_{,\tilde{\varphi}}\dfrac{1}{\pi_{\tilde{\alpha}}},\\
	\vartheta_e^q&=e^{-2\tilde{\alpha}}\mathcal{H}_0^{(2)}\left(19-18\dfrac{\mathcal{H}_0^{(2)}}{\pi_{\tilde{\alpha}}^2}\right)+e^{4\tilde{\alpha}}(\bar{W}_{,\tilde{\varphi}\tilde{\varphi}}-4\bar{W})\label{varthetaeq},\\
	\vartheta_T^q&=e^{-2\tilde{\alpha}}\mathcal{H}_0^{(2)}-4e^{4\tilde{\alpha}}\bar{W},\label{varthetaTq}
\end{align}
where $\bar{W}_{,\tilde{\varphi}}$ and $\bar{W}_{,\tilde{\varphi}\tilde{\varphi}}$ denote the first and second derivatives of the function $\bar{W}$ with respect to $\tilde{\varphi}$, respectively.

Inspecting the Hamiltonian \eqref{H}, one can realize that the relevant functions of the FLRW background for the scalar and tensor contributions are $(\vartheta_e^q+\vartheta_o\pi_{\tilde{\varphi}})/\vartheta_e$ and $\vartheta^q_T/\vartheta_e$, that differ in a quantity that vanishes with the field potential:
\begin{align}\label{relation}
\frac{\vartheta_e^q+\vartheta_o\pi_{\tilde{\varphi}}}{\vartheta_e}=\frac{\vartheta_T^q}{\vartheta_e}+e^{2\tilde{\alpha}}\left(\bar{W}_{,\tilde{\varphi}\tilde{\varphi}}+36\bar{W}\dfrac{\mathcal{H}_0^{(2)}}{\pi_{\tilde{\alpha}}^2}-12\bar{W}_{,\tilde{\varphi}}\dfrac{\pi_{\tilde{\varphi}}}{\pi_{\tilde{\alpha}}}\right).
\end{align}

Following the program of the hybrid approach, we select the polymeric representation of LQC for the FLRW geometry, so that the homogeneous gravitational degrees of freedom are encoded in a triad-like variable $V$, that corresponds to the physical volume of the Universe, and a canonically conjugate connection-like variable $b$ that, classically, is proportional to the Hubble parameter. Indeed, $\{b,V\}=4\pi G \gamma \sqrt{\Delta}$, where  $\gamma$ is the Immirzi parameter (that is usually assumed to take the value $\gamma_{\rm stand}\approx 0.2375$, derived from calculations of black-hole entropy) and $\Delta=4\sqrt{3}\pi G \gamma$ is called the area gap.  The relation of $V$ and $b$ with the logarithmic scale factor and its momentum is 
\begin{equation}
e^{\tilde{\alpha}}=\left(\dfrac{3}{4\pi G \sigma}\right)^{1/3}V^{1/3},\quad
\pi_{\tilde{\alpha}}=-\dfrac{3}{4\pi G \gamma \sqrt{\Delta}}bV . \label{pialpha}
\end{equation}
The regularization scheme adopted for the contribution of the homogeneous geometry in the Hamiltonian constraint leaves an imprint that is transmitted to the effective dynamics. In the present study, we consider the regularization resulting from the Dapor--Liegener prescription \cite{DL}, which is based on an individual treatment of the Euclidean and Lorentzian parts of the Hamiltonian. The component $\mathcal{H}_0^{(2)}$ of the densitized homogeneous Hamiltonian obtained with this prescription adopts the effective value \cite{DLhLQC}
\begin{align}
\mathcal{H}_0^{(2)}=-\left(\dfrac{3}{4\pi G}\right)^2V^2\left[\dfrac{1}{\Delta}\left(\sin^2b-\dfrac{1+\gamma^2}{4\gamma^2}\sin^22b\right)+2\dfrac{\bar{W}}{\sigma^2}\right].
\end{align}
Apart from powers of the volume, only two other geometrical quantities of the FLRW background remain to be defined in the densitized Hamiltonian constraint: the inverse of $\pi_{\tilde{\alpha}}$ and the square inverse, both in the scalar contribution [see Eqs. \eqref{varthetae} and \eqref{varthetaeq}]. Actually, the way in which these objects are represented is intimately related to the structure of the quantum theory: we require that their action as quantum operators preserve the superselection sectors of the homogeneous Hamiltonian \cite{GIper,DLhLQC}, so that the perturbative contributions do not alter them. In the first place, 
\begin{align}\label{1/pi}
\dfrac{1}{\pi_{\tilde{\alpha}}}=\dfrac{1}{\pi_{\tilde{\alpha}}^2}\,\Xi,
\end{align}
where $\Xi$ is the effective analog of the operator $\hat{\pi}_{\tilde{\alpha}}$. According to the proposal of Ref. \cite{GIper}{,}
\begin{align}\label{Xi}
\Xi=-\dfrac{3}{8\pi G\gamma\sqrt{\Delta}}V \sin 2b.
\end{align}
In this way, we are only left with the definition of $1/\pi_{\tilde{\alpha}}^2$. There exists an ambiguity in the quantum representation of this factor, that obviously affects its effective counterpart. In Ref. \cite{DLhLQC}, two different prescriptions were argued to be admissible for this representation, depending on whether or not one employs $\mathcal{H}_0^{(2)}$ in the construction. We will consider in detail these two prescriptions  in Secs. \ref{sec:introA} and \ref{sec:introB}. 

In the effective regime of homogeneous LQC, the mode equations that one can derive from the Hamiltonian constraint for the gauge invariant perturbations, can be cast in the form (see Refs. \cite{GIper,Olm})
\begin{align}
v_{\vec{n},\epsilon}''+\left(\tilde{\omega}_n^2+M^S\right)v_{\vec{n},\epsilon}&=0,\\
\tilde{d}_{\vec{n},\epsilon,\tilde{\epsilon}}''+\left(\tilde{\omega}_n^2+M^T\right)\tilde{d}_{\vec{n},\epsilon,\tilde{\epsilon}}&=0,
\end{align}
where the prime denotes the derivative with respect to conformal time, $\tilde{\omega}_n^2=l_0^2\omega_n^2$ and
\begin{align}
M^S=l_0^2\dfrac{\vartheta_e^q+\vartheta_o\pi_{\tilde{\varphi}}}{\vartheta_e},\quad
M^T=l_0^2\dfrac{\vartheta_T^q}{\vartheta_e},\label{MSdef}
\end{align}
that we will respectively call the \emph{scalar} and \emph{tensor effective masses}. By virtue of the relation \eqref{relation}, we immediately see that these two objects differ by a quantity that vanishes when the scalar field potential is identically zero, $M^S=M^T+\mathcal{U}$, where $\mathcal{U}$ is the so-called Mukhanov--Sasaki potential. Following the conventions of Ref. \cite{positividad}, we now reexpress these masses in terms of the matter energy density and pressure, defined in the standard way in terms of the matter Hamiltonian:
\begin{align}
\rho&=\dfrac{H_{\rm matt}}{V}=\dfrac{2\pi G}{3V^2}\left[\pi_{\tilde{\varphi}}^2+2\left(\dfrac{3}{4\pi G}\right)^2V^2\dfrac{\bar{W}}{\sigma^2}\right]=\dfrac{2\pi G }{3V^2}\pi_{\tilde{\alpha}}^2=\dfrac{3}{8\pi G\gamma^2 \Delta}\sin^2b[1-(1+\gamma^2)\sin^2b],\label{rho}\\
P&=-\dfrac{\partial H_{\rm matt}}{\partial V}=\rho-\dfrac{3}{2\pi G}\dfrac{\bar{W}}{\sigma^2}.\label{P}
\end{align}
In the two last equalities of Eq. \eqref{rho}, we have used that, at our perturbative truncation order, we can ignore the backreaction of the perturbations in the computation of the masses, and then we have employed the vanishing of the Hamiltonian constraint of effective homogeneous LQC.

On the one hand, the tensor effective mass can be rewritten in terms of $\rho$ and $P$ as
\begin{align}
M^T=V^{-4/3}\left[\left(\dfrac{4\pi G}{3}\right)^2\pi_{\tilde{\alpha}}^2-6V^2\dfrac{\bar{W}}{\sigma^2}\right]=-\dfrac{4\pi G}{3}V^{2/3}(\rho-3P). \label{exprMT}
\end{align}
On the other hand, the Mukhanov--Sasaki potential $\mathcal{U}$ is [see Eqs. \eqref{relation} and \eqref{MSdef}]
\begin{align}
\mathcal{U}=\dfrac{V^{2/3}}{\sigma^2}\left[\bar{W}_{,\tilde{\varphi}\tilde{\varphi}}+72\dfrac{3}{4\pi G}\bar{W}\dfrac{V^2\rho}{\pi_{\tilde{\alpha}}^2}-12\bar{W}_{,\tilde{\varphi}}\dfrac{\Xi\pi_{\tilde{\varphi}}}{\pi_{\tilde{\alpha}}^2}-72\left(\dfrac{3}{4\pi G \sigma}\right)^2\bar{W}^2\dfrac{V^2}{\pi_{\tilde{\alpha}}^2}\right],\label{U}
\end{align}
where only $\Xi$ [see Eq. \eqref{Xi}] and $1/\pi_{\tilde{\alpha}}^2$ remain to be determined in terms of $\rho$ and $P$. This task can only be done after having selected a concrete representation for $1/\pi_{\tilde{\alpha}}^2$. The computation is performed individually for each of the two considered prescriptions in Secs. \ref{sec:introA} and \ref{sec:introB}. Notice that, since the tensor effective mass does not depend on inverse powers of the momentum of the logarithmic scale factor, its effective value is the same for both prescriptions. Only the scalar effective mass or, equivalently, the Mukhanov--Sasaki potential, needs to be studied separately.

\subsection{Prescription A}\label{sec:introA}
The first option to regularize the square inverse of the momentum of the logarithmic scale factor is naturally provided by the term $\mathcal{H}_0^{(2)}$ that contains the geometrical part of the densitized homogeneous Hamiltonian. Indeed, from Eq. \eqref{H02}, we can define $1/\pi_{\tilde{\alpha}}^2$ as
\begin{align}
\dfrac{1}{\pi_{\tilde{\alpha}}^2}=\left[\mathcal{H}_0^{(2)}+2\left(\dfrac{3}{4\pi G}\right)^2V^2\dfrac{\bar{W}}{\sigma^2}\right]^{-1}=-\left(\dfrac{4\pi G \sqrt{\Delta}}{3}\right)^2\dfrac{1}{V^2}\left(\sin^2b-\dfrac{1+\gamma^2}{4\gamma^2}\sin^22b\right)^{-1}=\dfrac{2\pi G}{3}\dfrac{1}{V^2\rho},
\end{align}
where in the second equality we have substituted the effective expression of the considered quantity in the Dapor--Liegener proposal, and in the last equality we have used the vanishing of the effective homogeneous constraint, since we can ignore the backreaction at the order of our approximations in the mass. Throughout this paper, we refer to this way of representing $1/\pi_{\tilde{\alpha}}^2$ and the subsequent effective value as ``prescription A''. Together with Eqs. \eqref{1/pi} and \eqref{Xi}, this implies that
\begin{align}
\dfrac{\Xi}{\pi_{\tilde{\alpha}}^2}=\dfrac{2\pi G \sqrt{\Delta}}{3\gamma}\dfrac{\sin 2b}{V}\left(\sin^2b-\dfrac{1+\gamma^2}{4\gamma^2}\sin^22b\right)^{-1}=-\dfrac{1}{4\gamma\sqrt{\Delta}}\dfrac{\sin 2b}{V\rho}.
\end{align}

We can then write the Mukhanov--Sasaki potential in the form
\begin{align}
{}^{(A)}\mathcal{U}=\dfrac{V^{2/3}}{\sigma^2}\left(\bar{W}_{,\tilde{\varphi}\tilde{\varphi}}+36\bar{W}+\dfrac{3}{\gamma\sqrt{\Delta}}\sin 2b\dfrac{\bar{W}_{,\tilde{\varphi}}\pi_{\tilde{\varphi}}}{V\rho}-36\dfrac{3}{4\pi G \sigma^2}\dfrac{\bar{W}^2}{\rho}\right),\label{UA}
\end{align}
where $\sin 2b$ would need to be rewritten in terms of $\rho$. This cannot be done globally. Indeed, Eq. \eqref{rho} can be used to relate the energy density  with $\sin^2b$. The resulting equation admits two solutions,
\begin{align}\label{sin2brho}
\sin^2b_{\pm}=\dfrac{1\pm\sqrt{1-\rho/\rho_B}}{2(1+\gamma^2)},
\end{align}
where $\rho_B$ is the maximum of the energy density, attained at the bounce:
\begin{align}
\rho_B=\dfrac{3}{32\pi G\gamma^2(1+\gamma^2)\Delta}.
\end{align}
The existence of two solutions leads to two different branches in the evolution, that cannot be joined arbitrarily. There are only two possible evolution histories: an asymmetric bounce may join an asymptotically de Sitter universe with a flat FLRW cosmology or the other way around \cite{DL,Paramc1}. In this instance, we work with the solution that is not immediately ruled out by cosmological observations: a contracting asymptotically de Sitter branch followed by an expanding flat FLRW universe. Then, in Eq. \eqref{sin2brho}, the plus sign corresponds to the contracting branch, while the minus sign is assigned to the postbounce evolution. By substituting relation \eqref{sin2brho} in Eq. \eqref{UA}, we get two expressions for the Mukhanov--Sasaki potential that differ in this sign, each of them valid before or after the bounce.

\subsection{Prescription B}\label{sec:introB}
Let us consider now the second prescription for the representation of $1/\pi_{\tilde{\alpha}}^2$. Although admissible, this possibility, that we call ``prescription B'', lacks the motivation that is the theoretical strength of prescription A. Following the proposal of Ref. \cite{GIper}, we can use the same representation of $1/\pi_{\tilde{\alpha}}^2$ as in the standard regularization of LQC, therefore mixing regularizations in a certain sense. This leads to the effective value 
\begin{align}\label{presB}
\dfrac{1}{\pi_{\tilde{\alpha}}^2}=\left(\dfrac{4\pi G \gamma \sqrt{\Delta}}{3}\right)^2\dfrac{1}{V^2\sin^2b}.
\end{align}
Details about the quantum features of this prescription can be found in Ref. \cite{DLhLQC}. In this manner, we obtain
\begin{align}
\dfrac{\Xi}{\pi_{\tilde{\alpha}}^2}=-\dfrac{4\pi G\gamma \sqrt{\Delta}}{3V}\dfrac{\cos b}{\sin b},
\end{align}
and, hence, the following effective expression of the Mukhanov--Sasaki potential is found: 
\begin{align}
{}^{(B)}\mathcal{U}=\dfrac{V^{2/3}}{\sigma^2}\left(\bar{W}_{,\tilde{\varphi}\tilde{\varphi}}+96\pi G \gamma^2\Delta\dfrac{\bar{W}\rho}{\sin^2b}+16\pi G \gamma\sqrt{\Delta}\,\dfrac{\cos b}{\sin b}\dfrac{\bar{W}_{,\tilde{\varphi}}\pi_{\tilde{\varphi}}}{V}-\dfrac{72\gamma^2\Delta}{\sigma^2}\dfrac{\bar{W}^2}{\sin^2b}\right).
\label{UB}
\end{align}
Employing Eq. \eqref{sin2brho}, we can finally rewrite the trigonometric functions of $b$ in terms of the matter energy density, although, as for prescription A, this cannot be done globally, but separately in the prebounce and postbounce branches. 

One of our objectives is to compare prescriptions A and B so as to verify whether the theoretical preference for prescription A also translates into more appealing physical features for the corresponding masses. This analysis is carried out in Secs. \ref{sec:bounce} and \ref{sec:dS}, where we also consider the effective value of the masses in the dressed metric approach.

\section{The dressed metric approach}\label{sec:dressed}

In this section, we discuss the other predominant approach to inhomogeneous LQC that is also based on a combination of different quantum representations for the homogeneous and inhomogeneous sectors of a perturbed cosmology: the dressed metric approach. We will obtain explicitly the effective value of the time-dependent masses that govern the dynamics of the scalar and tensor perturbations. For specific details and in-depth discussions about the dressed metric approach, we refer the reader to Refs. \cite{dressed1,dressed2,dressed3,dressed4}.

This approach to inhomogeneous LQC adopts a separate treatment of the background homogeneous cosmology and the perturbations, neglecting the effects of the backreaction from the very beginning. One deals with the homogeneous sector first, obtaining an FLRW cosmology that is ``dressed'' with the main quantum corrections in a sort of mean field approximation.  The homogeneous dynamics of this dressed cosmology is lifted to the truncated phase space that aims to describe the perturbed cosmology at the desired level of approximation \cite{dressed2}. In this way, the perturbations can be seen as test fields propagating on a dressed homogeneous background. It is important to note that, as a result of this program, one is devoid of a global Hamiltonian. In fact, the dressed metric formalism is provided with two different Hamiltonians, one of them generating the homogeneous dynamics and the other one leading to the field equations for the gauge invariant perturbations once the dressed background is viewed as a given entity \cite{dressed2,dressed3}.

Employing the notation of Refs. \cite{dressed2,dressed3} and the transformations of variables considered in Ref. \cite{positividad} (that facilitate a direct comparison with the hybrid approach), the dynamical equations for the perturbations in the effective regime of LQC can be cast in the form
\begin{align}
q_{\vec{n},\epsilon}''+[\tilde{\omega}_n^2+{}^{(D)}M^S]\,q_{\vec{n},\epsilon}&=0,\\
t_{\vec{n},\epsilon,\tilde{\epsilon}}''+[\tilde{\omega}_n^2+{}^{(D)}M^T]\,t_{\vec{n},\epsilon,\tilde{\epsilon}}&=0,
\end{align}
where $q_{\vec{n},\epsilon}$ and $t_{\vec{n},\epsilon,\tilde{\epsilon}}$ are the corresponding Mukhanov--Sasaki and tensor modes, respectively. We have introduced a  notation slightly different to that used in the discussion of the hybrid approach in order to distinguish between the variables of the two formalisms, since they do not obey exactly the same equations. The scalar and tensor effective masses in the dressed metric formalism are given by
\begin{align}
{}^{(D)}M^T&=-\dfrac{(e^{\tilde{\alpha}})''}{e^{\tilde{\alpha}}}=-\dfrac{V''}{3V}+2\left(\dfrac{V'}{3V}\right)^2,\\
{}^{(D)}M^S&={}^{(D)}M^T+{}^{(D)}\mathcal{U},
\end{align}
where ${}^{(D)}\mathcal{U}$ is the Mukhanov--Sasaki potential in the dressed metric approach.

Let us begin by obtaining the expression of the tensor effective mass in terms of the energy density and pressure of the scalar field. In order to compute the derivative of any function of the homogeneous phase space with respect to the conformal time, in the dressed metric approach we simply have to take Poisson brackets with the effective homogeneous Hamiltonian that generates the evolution on the trajectories described by the dressed background solution. For instance, in the case under consideration,
\begin{align}\label{V'}
V'&=-\dfrac{2\pi G}{3}V^{-2/3}\{V,\mathcal{H}_0^{(2)}\}=-\dfrac{3\gamma}{2\sqrt{\Delta}}V^{4/3}\sin 2b \left(1-\dfrac{1+\gamma^2}{\gamma^2}\cos 2b\right).
\end{align}
Similarly,
\begin{align}\label{V''}
V''&=-\dfrac{3\gamma^2}{\Delta}V^{5/3}\sin^2b\left\{3\left[\cos 2b-\dfrac{1+\gamma^2}{\gamma^2}(\cos^2 2b-\sin^2 2b)\right]\left(1-\dfrac{1+\gamma^2}{\gamma^2}\cos^2b\right)\right.  \nonumber\\
&-\left.4\cos^2b\left(1-\dfrac{1+\gamma^2}{\gamma^2}\cos 2b\right)^2\right\}-18\gamma^2V^{5/3}\dfrac{\bar{W}}{\sigma^2}\left[\cos 2b-\dfrac{1+\gamma^2}{\gamma^2}(\cos^2 2b-\sin^2 2b)\right].
\end{align}
In conclusion, the effective mass for the tensor modes in the dressed metric approach is
\begin{align}\label{MTD}
{}^{(D)}M^T&=\dfrac{\gamma^2}{\Delta}V^{2/3}\sin^2b\left\{3\left[\cos 2b-\dfrac{1+\gamma^2}{\gamma^2}(\cos^2 2b-\sin^2 2b)\right]\left(1-\dfrac{1+\gamma^2}{\gamma^2}\cos^2b\right)\right.  \nonumber\\
&-\left.2\cos^2b\left(1-\dfrac{1+\gamma^2}{\gamma^2}\cos 2b\right)^2\right\}+6\gamma^2V^{2/3}\dfrac{\bar{W}}{\sigma^2}\left[\cos 2b-\dfrac{1+\gamma^2}{\gamma^2}(\cos^2 2b-\sin^2 2b)\right].
\end{align}
This is a complicated expression of the holonomy elements of the connection-like variable $b$, which could in principle be reexpressed in terms of $\rho$ and $P$. In this process, we need to use Eq. \eqref{sin2brho}, resulting in two different analytical expressions of the tensor effective mass if one wants to cover the entire evolution of the Universe.  It is straightforward to find that
\begin{align}
\cos 2b_{\pm}-\dfrac{1+\gamma^2}{\gamma^2}(\cos^22b_\pm-\sin^22b_\pm)&=-\dfrac{1}{\gamma^2}+\dfrac{3}{1+\gamma^2}\left(1\pm\sqrt{1-\dfrac{\rho}{\rho_B}}\right)+\dfrac{2}{\gamma^2(1+\gamma^2)}\dfrac{\rho}{\rho_B},\label{dS1}\\
1-\dfrac{1+\gamma^2}{\gamma^2}\cos^2b_\pm&=-\dfrac{1}{2\gamma^2}\left(1\mp\sqrt{1-\dfrac{\rho}{\rho_B}}\right),\label{dS2}\\
1-\dfrac{1+\gamma^2}{\gamma^2}\cos 2b_{\pm}&=\pm\dfrac{1}{\gamma^2}\sqrt{1-\dfrac{\rho}{\rho_B}}\label{dS3},\\
\dfrac{\bar{W}}{\sigma^2}&=\dfrac{2\pi G}{3}(\rho-P)\label{dS4},
\end{align}
where we recall that the plus and minus signs correspond to the asymptotically de Sitter branch and the expanding FLRW cosmology, respectively. With these equations, it is trivial to rewrite ${}^{(D)}M^T$ in the desired form.

At this point of our analysis, only the scalar effective mass remains to be discussed. As in the hybrid approach, this mass is the sum of the tensor effective mass and the Mukhanov--Sasaki potential. Therefore, it is sufficient to discuss the expression of ${}^{(D)}\mathcal{U}$. In practice, the Mukhanov--Sasaki potential can be found from Eq. \eqref{U} by simply substituting in it the effective value of the momentum of the logarithmic scale factor, expressed in terms of the matter energy density and the scalar field potential. Imposing the effective homogeneous Hamiltonian as a constraint, because the backreaction is ignored, the expression that results from this procedure is
\begin{align}\label{UD}
{}^{(D)}\mathcal{U}=\dfrac{V^{2/3}}{\sigma^2}\left(\bar{W}_{,\tilde{\varphi}\tilde{\varphi}}+36\bar{W}-s\, 4\sqrt{6\pi G}\dfrac{|\pi_{\tilde{\varphi}}|}{V}\dfrac{\bar{W}_{,\tilde{\varphi}}}{\sqrt{\rho}}-36\dfrac{3}{4\pi G \sigma^2}\dfrac{\bar{W}^2}{\rho}\right),
\end{align}
where $s$ is the sign of $\pi_{\tilde{\varphi}}/\pi_{\tilde{\alpha}}$. It is interesting to note that this formula for the Mukhanov--Sasaki potential coincides with that found for prescription A in the hybrid approach [see Eq. \eqref{UA}] when $s$ is replaced with $-3\,\rm{sgn}(\pi_{\tilde{\varphi}})\sin 2b/(4\gamma\sqrt{6\pi G\Delta\rho})$.

Once we have specified the effective masses of the perturbations in the hybrid and dressed metric approaches, we are in an adequate position to analyze their properties in regimes of physical interest and compare the results corresponding to each approach. This is the objective of the next two sections.

\section{Effective masses at the bounce}\label{sec:bounce}

We now want to analyze the properties at the bounce of the effective masses introduced above for prescriptions A and B within the hybrid approach{,} and for the dressed metric approach. We are interested in elucidating the positivity of these masses and, with this aim, we will carry out a study similar to the one presented in Ref. \cite{positividad} in the context of the standard regularization scheme for homogeneous LQC.

Given that the bounce is characterized by the fact that the physical volume $V$ reaches a minimum, it is straightforward to realize that Eq. \eqref{V'} [with Eq. \eqref{V''}] implies that, at the bounce,
\begin{align}
\sin^2b_B=\dfrac{1}{2(1+\gamma^2)},
\end{align}
where the subindex (or superindex) $B$ stands for evaluation at the instant of the bounce. In the Dapor--Liegener model, the solution of cosmological interest is the one where the asymptotically de Sitter branch appears in the prebounce era, for which one only has to consider positive values of $b$ \cite{Paramc1,Haro}. Thus, we set
\begin{align}\label{bB}
b_B= \sin^{-1} \left(\dfrac{1}{\sqrt{2(1+\gamma^2)}}\right),
\end{align}
so that we select the smallest positive value allowed for $b_B$. In the discussed solution, it is found that $b$ is a monotonically decreasing function of the proper time, that attains its maximum in the infinite past and decreases to zero in the infinite future \cite{Paramc1,Haro}.  Let us now evaluate at the bounce the effective masses that we have defined.

\subsection{Prescription A}\label{sec:bounceA}

According to Eq. \eqref{exprMT}, the tensor effective mass in the hybrid approach, that it is independent of the prescription A or B chosen to represent the square inverse of the momentum of the logarithmic scale factor, adopts the following value at the bounce:
\begin{align}\label{MThybridB}
M^T_B=-\dfrac{4\pi G}{3}V_B^{2/3}(\rho_B-3P_B)=\dfrac{8\pi G}{3}V_B^{2/3}\left(\rho_B-\dfrac{9}{4\pi G \sigma^2}\bar{W}_B\right).
\end{align}
As a result, this mass is positive  at the bounce as long as
\begin{align}
\dfrac{3}{4\pi G \sigma^2}\bar{W}_B<\dfrac{\rho_B}{3}.
\end{align}
Notice that this upper bound (and, thus, the positivity of the effective mass seen by the tensor perturbations) is satisfied in a scenario of kinetic dominance at the bounce. Indeed, the bound is equivalent to the condition that the kinetic contribution to the energy density be greater than $2\rho_B/3$ and, therefore, at least twice as large as the potential contribution. So, in conclusion, when the energy density of the scalar field is kinetically dominated at the bounce, the tensor effective mass is strictly positive in the hybrid approach. This conclusion is identical to the one that was reached with the standard regularization scheme in LQC and, hence, it is thereby provided of robustness.

Let us now analyze the scalar effective mass. Using Eq. \eqref{UA}, we obtain 
\begin{align}\label{MBSA1}
{}^{(A)}M_B^S=\dfrac{8\pi G}{3}V_B^{2/3}\left[\rho_B+\dfrac{3}{8\pi G \sigma^2}\left(\bar{W}^B_{,\tilde{\varphi}\tilde{\varphi}}+30\bar{W}_B+\dfrac{3}{\gamma\sqrt{\Delta}}\sin 2b_B\dfrac{\bar{W}^B_{,\tilde{\varphi}}\pi^B_{\tilde{\varphi}}}{V_B\rho_B}-36\dfrac{3}{4\pi G \sigma^2}\dfrac{\bar{W}_B^2}{\rho_B}\right)\right],
\end{align}
where
\begin{align}
\sin 2b_B=\dfrac{\sqrt{1+2\gamma^2}}{1+\gamma^2}.
\end{align}

In order to be able to study the properties of this mass analytically, we need to introduce some restrictions on the scalar field potential, that we will motivate so as to cover the case of a mass term, in which we are particularly interested. In the following, we particularize to scalar field potentials that verify {\emph{at the bounce}} that
\begin{align}
\bar{W}_B\geq 0,\qquad \bar{W}^B_{,\tilde{\varphi}\tilde{\varphi}}\geq 0.\label{cond1}
\end{align}
Because of the presence of a term proportional to the first derivative of the potential with respect to the field in Eq. \eqref{MBSA1}, we still  restrict our attention to potentials such that, {\emph{at the bounce}},
\begin{align}
|\bar{W}^B_{,\tilde{\varphi}}|\leq C\sqrt{2\bar{W}_B\bar{W}^B_{,\tilde{\varphi}\tilde{\varphi}}},\label{cond2}
\end{align}
with $C=C(\bar{W}^B_{,\tilde{\varphi}\tilde{\varphi}})$ being a positive function of order one. It is noteworthy that the relevant case of a mass term that we have mentioned belongs to the family of potentials that satisfy conditions \eqref{cond1} and \eqref{cond2}. Indeed, the mass potential displays the properties \eqref{cond1} at all times and satisfies the identity $|\bar{W}_{,\tilde{\varphi}}|=(2\bar{W}\bar{W}_{,\tilde{\varphi}\tilde{\varphi}})^{1/2}$. Hence, in this case, the function $C$ can be made equal to the unit. We also notice that the introduced conditions on the scalar field potential are the same that were considered at the bounce in the study of the effective masses for the standard regularization scheme in Ref. \cite{positividad}.

Employing the definition of the matter energy density \eqref{rho} at the bounce, one can show that \cite{positividad}
\begin{align}
\dfrac{|\pi_{\tilde{\varphi}}^B|\sqrt{2\bar{W}_B}}{V_B}=\sqrt{\dfrac{3}{\pi G}\bar{W}_B\left(\rho_B-\dfrac{3}{4\pi G\sigma^2}\bar{W}_B\right)}.\label{maxexpr}
\end{align}
The right-hand side of this equation, understood as a function of $\bar{W}_B$, has a local maximum and its value at that point provides the bound ${|\pi_{\tilde{\varphi}}^B|\sqrt{2\bar{W}_B}}/{V_B}\leq\sigma \rho_B$. Combining this upper bound with condition \eqref{cond2} on the first derivative of the scalar field potential, we find that the absolute value of the term proportional to $\bar{W}^B_{,\tilde{\varphi}}$ in Eq. \eqref{MBSA1} satisfies
\begin{align}
\dfrac{|\bar{W}^B_{,\tilde{\varphi}}\pi_{\tilde{\varphi}}^B|}{V_B\rho_B}\leq \sigma C\sqrt{\bar{W}^B_{,\tilde{\varphi}\tilde{\varphi}}}.
\end{align}
This upper bound, together with the fact that the derivative of the potential can adopt any sign, then allows us to bound the scalar effective mass at the bounce both from above and below. Indeed, we have $P_-\leq {}^{(A)}M_B^S\leq P_+$ where 
\begin{align}
P_{\pm}=\dfrac{8\pi G}{3}V_B^{2/3}\left[\rho_B+\dfrac{3}{8\pi G \sigma^2}\left(\bar{W}^B_{,\tilde{\varphi}\tilde{\varphi}}+30\bar{W}_B\pm\dfrac{3\sigma C}{\gamma \sqrt{\Delta}}\dfrac{\sqrt{1+2\gamma^2}}{1+\gamma^2}\sqrt{\bar{W}^B_{,\tilde{\varphi}\tilde{\varphi}}}-36\dfrac{3}{4\pi G \sigma^2}\dfrac{\bar{W}_B^2}{\rho_B}\right)\right],
\end{align}
are regarded as quadratic polynomials in $\bar{W}_B$. The roots of these polynomials, denoted by $x_\pm(P_+)$ and $x_{\pm}(P_-)$, are 
\begin{align}\label{xP}
\dfrac{3}{4\pi G \sigma^2}x_\pm(P_\pm)=\dfrac{5\pm \sqrt{33+\dfrac{3}{\pi G\sigma^2}\dfrac{\bar{W}^B_{,\tilde{\varphi}\tilde{\varphi}}}{\rho_B}\pm \dfrac{9C}{\pi G \sigma \gamma \sqrt{\Delta}}\dfrac{\sqrt{1+2\gamma^2}}{1+\gamma^2}\dfrac{\sqrt{\bar{W}^B_{,\tilde{\varphi}\tilde{\varphi}}}}{\rho_B}}}{12}\rho_B,
\end{align}
where the $\pm$ sign inside the square root corresponds to $P_{\pm}$, respectively.

For a scalar field potential of the class we are considering, with a nonnegative second derivative at the bounce, both of the roots of $P_+$ are real. Therefore, the upper bound ${}^{(A)}M_B^S\leq P_+$ entails that the scalar effective mass at the bounce is ensured to be negative when the same happens to $P_+$ or, equivalently, if
\begin{align}
\bar{W}_B\in \left[0,\dfrac{4\pi G \sigma^2}{3}\rho_B\right]\setminus \left([x_-(P_+),x_+(P_+)]\cap\left[0,\dfrac{4\pi G \sigma^2}{3}\rho_B\right]\right),
\end{align}
where we have used that, since the kinetic contribution to the energy density is always nonnegative, the maximum value allowed for $\bar{W}$ at the bounce is $4\pi G \sigma^2\rho_B/3$. Given that $x_-(P_+)$ is trivially negative, we conclude that the scalar effective mass is negative in the region defined by
\begin{align}
\bar{W}_B\in \left(x_+(P_+),\dfrac{4\pi G\sigma^2}{3}\rho_B\right],
\end{align}
provided that $x_+(P_+)<4\pi G \sigma^2\rho_B/3$. Otherwise, the upper bound of the scalar effective mass would not provide any information about the region of physical potentials where ${}^{(A)}M_B^S$ is ensured to be negative. 

It is straightforward so see that $x_+(P_+)<4\pi G \sigma^2\rho_B/3$ if and only if
\begin{align}
w^2+4\sqrt{6}C\sqrt{\dfrac{1+2\gamma^2}{1+\gamma^2}}w-\dfrac{16}{3}<0,
\end{align}
where we have introduced the notation $w= (\bar{W}^B_{,\tilde{\varphi}\tilde{\varphi}})^{1/2}/(\pi G \sigma^2 \rho_B)^{1/2}\in[0,\infty^+)$. This inequality is satisfied when $w\in (w_-^{(A)},w_+^{(A)})\cap [0,\infty^+)$, where $w_\pm^{(A)}$ are the roots of the polynomial on the left-hand side of the inequality:
\begin{align}
w_\pm^{(A)}=-2\sqrt{6}C\sqrt{\dfrac{1+2\gamma^2}{1+\gamma^2}}\pm \sqrt{24C^2\dfrac{1+2\gamma^2}{1+\gamma^2}+\dfrac{16}{3}}.
\end{align}
Since $w_{+}^{(A)}$ and $w_-^{(A)}$ are obviously positive and negative, respectively, there exists a region of the space of field potentials where the scalar effective mass at the bounce is ensured to be negative if and only if the second derivative of the field potential at that instant (that we have restricted already to be nonnegative) is small enough:
\begin{align}
\dfrac{\bar{W}^B_{,\tilde{\varphi}\tilde{\varphi}}}{\pi G \sigma^2\rho_B}\in\left[0,\left(w_+^{(A)}\right)^2\right).\label{wbb}
\end{align}
Taking $C=1$ and the standard value of the Immirzi parameter, we obtain $w_+^{(A)}\approx 0.505$.

Restriction \eqref{wbb} allows for values of $\bar{W}^B_{,\tilde{\varphi}\tilde{\varphi}}$ in a neighborhood of zero, which is an especially interesting case. The region where we can ensure that ${}^{(A)}M_B^S$ is negative gets larger as $\bar{W}^B_{,\tilde{\varphi}\tilde{\varphi}}$ goes to zero, so that its maximum extension is reached when the second derivative of the scalar field potential at the bounce vanishes. In this case, the scalar effective mass at the bounce is negative when
\begin{align}\label{negA}
\dfrac{3}{4\pi G \sigma^2\rho_B}\bar{W}_B\in\left(\dfrac{5+\sqrt{33}}{12},1\right],
\end{align}
which roughly represents a $10.5\%$ of the space of physical potentials. 

On the other hand, if the roots of $P_-$ are real as well, something that certainly happens for values of $\bar{W}^B_{,\tilde{\varphi}\tilde{\varphi}}$ in a neighborhood of zero, the lower bound ${}^{(A)}M_B^S\geq P_-$ guarantees that the scalar effective mass at the bounce is nonnegative when $\bar{W}_B\in [x_-(P_-),x_+(P_-)]\cap [0,4\pi G \sigma^2\rho_B/3]$. Let $R_-$ be the argument of the square root in Eq. \eqref{xP} corresponding to $P_-$. Then, the roots $x_\pm(P_-)$ are complex if $R_-<D$ for $D=0$. The smaller root $x_-(P_-)$ is positive provided that $R_-<D$ for $D=25$. The larger one $x_+(P_-)$ is smaller than $4\pi G \sigma^2\rho_B/3$ as long as $R_-<D$ for $D=49$. As a result, depending on the value of $D$ with respect to these transitional ones, four distinct situations are possible: (i) if $D<0$, both roots are complex and no new information comes to light; (ii) if $0\leq D<25$, ${}^{(A)}M_B^S$ is nonnegative when $\bar{W}_B\in [x_-(P_-),x_+(P_-)]$; (iii) if $25\leq D < 49$, ${}^{(A)}M_B^S$ is nonnegative when $\bar{W}_B\in [0,x_+(P_-)]$; and (iv) if $D\geq 49$, ${}^{(A)}M_B^S$ is nonnegative for all $\bar{W}_B\in[0,4\pi G \sigma^2\rho_B/3]$. 

Recasting the equality $R_-=D$ as a polynomial equation, quadratic in the variable $w=(\bar{W}^B_{,\tilde{\varphi}\tilde{\varphi}})^{1/2}/(\pi G \sigma^2\rho_B)^{1/2}$, that by construction is restricted to nonnegative values, it is straightforward to see for which values of the second derivative of the field potential each of the situations (i)-(iv) arises. Let $w_\pm (D)$ be the zeros of that polynomial, with the restriction to the positive semiaxis yet to be imposed. In the first place, using the definition \eqref{rho} of the matter energy density, one can verify that the polynomial does not have any real roots for $D<[33-72C^2(1+2\gamma^2)/(1+\gamma^2)]$. The quantity on the right-hand side is approximately $-42.8$ for $C=1$ and the standard value of the Immirzi parameter. Then, if the real values of $C$ and $\gamma$ lie on a certain neighborhood of the more natural ones, the quadratic polynomial in $w$ has two real roots for any $D\geq 0$. It is immediate to realize that the smaller root is not negative if and only if $D\leq 33$. Furthermore, for large $w$ the polynomial is always positive. Hence, we conclude that, for the standard values of $C$ and $\gamma$ or values close enough, the cases (i)-(iv) mentioned above take place when: (i) $w\in(w_-(0),w_+(0))$, (ii) $w\in(w_-(25),w_-(0)]\cup [w_+(0),w_+(25))$, (iii) $w\in[0,w_-(25)]\cup [w_+(25),w_+(49))$, and (iv) $w\geq w_+(49)$.

In the particular case where the scalar field potential is simply given by a mass term, its second derivative is constant and the value of $w$ is fixed by the mass of the scalar field alone. In the situations of interest for the phenomenology of the CMB, this mass is considerably small (see Refs. \cite{CMBhybrid,CMBdressed} for discussions within the context of both the hybrid and dressed metric formalisms) and, as a result, so is $w$. Therefore, these scenarios belong to the situation (iii). Then, there exists a region of the space of physical potentials at the bounce that contains $\bar{W}_B=0$ where the scalar effective mass is nonnegative at the bounce. For the aforementioned small values of the mass of the scalar field, this region extends up to field potentials close to $(5+\sqrt{33})\pi G \sigma^2\rho_B/9$ [namely, the value of $x_+(P_-)$ for $\bar{W}^B_{,\tilde{\varphi}\tilde{\varphi}}=0$], that covers the sector of solutions where the matter energy density at the bounce is kinetically dominated. 

In conclusion, using prescription A, the resulting scalar and tensor effective masses at the bounce are inevitably positive in the case of interest where the scalar field potential and its second derivative are small, found in the kinetically dominated scenarios at the bounce that lead to good fits of the observed CMB spectra, while still allowing for the presence of quantum effects at low multipoles.

\subsection{Prescription B}\label{sec:bounceB}

Let us explore now the consequences of adopting a representation of the square inverse of the momentum of the logarithmic scale factor as in Ref. \cite{GIper}, instead of employing the natural representation provided by the geometrical part of the (densitized) homogeneous scalar constraint in the Dapor--Liegener regularization scheme. The analysis is very similar to the one presented for prescription A. For this reason, we only point out the differences with respect to the previous case and write down the results for comparison.

In the first place, since the tensor effective mass is independent of the chosen representation for $1/\pi_{\tilde{\alpha}}^2$, its value at the bounce is also given by Eq. \eqref{MThybridB} and the comments on its positivity still hold.  On the other hand, as far as the scalar effective mass is concerned, it is immediate to derive from Eqs. \eqref{MThybridB} and \eqref{UB} that, at the bounce,
\begin{align}
{}^{(B)}M_B^S=\dfrac{8\pi G}{3}V_B^{2/3}\left[\rho_B+\dfrac{3}{8\pi G\sigma^2}\left(\bar{W}^B_{,\tilde{\varphi}\tilde{\varphi}}+12\bar{W}_B+16\pi G \gamma \sqrt{1+2\gamma^2} \sqrt{\Delta}\dfrac{\bar{W}^B_{,\tilde{\varphi}}\pi^B_{\tilde{\varphi}}}{V_B}-\dfrac{144\gamma^2(1+\gamma^2)\Delta}{\sigma^2}\bar{W}^2_B\right)\right].
\end{align}

Restricting our discussion to the scalar field potentials that satisfy Eqs. \eqref{cond1} and \eqref{cond2}, as in the previous subsection, we conclude that the scalar effective mass at the bounce is bounded above and below, as was the case in prescription A, by two polynomials quadratic in $\bar{W}_B$, namely $Q_-\leq{}^{(B)}M_B^S\leq Q_+$, where
\begin{align}
Q_\pm=\dfrac{8\pi G}{3}V_B^{2/3}\left[\rho_B+\dfrac{3}{8\pi G\sigma^2}\left(\bar{W}^B_{,\tilde{\varphi}\tilde{\varphi}}+12\bar{W}_B\pm \dfrac{3\sigma C}{2\gamma\sqrt{\Delta}}\dfrac{\sqrt{1+2\gamma^2}}{1+\gamma^2}\sqrt{\bar{W}^B_{,\tilde{\varphi}\tilde{\varphi}}}-\dfrac{144\gamma^2(1+\gamma^2)\Delta}{\sigma^2}\bar{W}_B^2\right)\right].
\end{align} 
The roots of these polynomials, $x_\pm(Q_{+})$ and $x_\pm(Q_-)$, are given by
\begin{align}\label{xQ}
\dfrac{3}{4\pi G \sigma^2}x_\pm(Q_\pm)=\dfrac{1\pm\sqrt{2+\dfrac{3}{8\pi G \sigma^2}\dfrac{\bar{W}^B_{,\tilde{\varphi}\tilde{\varphi}}}{\rho_B}\pm \dfrac{9C}{16\pi G \sigma\gamma\sqrt{\Delta}}\dfrac{\sqrt{1+2\gamma^2}}{1+\gamma^2}\dfrac{\sqrt{\bar{W}^B_{,\tilde{\varphi}\tilde{\varphi}}}}{\rho_B}}}{3}\rho_B,
\end{align}
with the plus sign inside the square root corresponding to $Q_+$. Given our restriction to $\bar{W}^B_{,\tilde{\varphi}\tilde{\varphi}}\geq 0$, both roots of $Q_+$ are real. As a result, and since $x_-(Q_+)$ is negative, the upper bound on ${}^{(B)}M_B^S$ implies that the scalar effective mass at the bounce is necessarily negative when the field potential takes values in the interval
\begin{align}\label{interx}
\left(x_+(Q_+),\dfrac{4\pi G \sigma^2}{3}\rho_B\right],
\end{align}
provided that $x_+(Q_+)<4\pi G \sigma^2\rho_B/3$. This condition is clearly satisfied for a sufficiently small $\bar{W}^B_{,\tilde{\varphi}\tilde{\varphi}}$. This requirement for the existence of a nonempty set of values of $\bar{W}_B$ for which the scalar effective mass at the bounce can be ensured to be negative is satisfied if and only if the second derivative of the scalar field potential is such that
\begin{align}\label{inwbb}
\dfrac{\bar{W}^B_{,\tilde{\varphi}\tilde{\varphi}}}{\pi G \sigma^2 \rho_B}\in \left[0,\left(w_+^{(B)}\right)^2\right),
\end{align}
with
\begin{align}\label{w+B}
w_+^{(B)}=-\sqrt{6}C\sqrt{\dfrac{1+2\gamma^2}{1+\gamma^2}}+\sqrt{6C^2\dfrac{1+2\gamma^2}{1+\gamma^2}+\dfrac{16}{3}}.
\end{align}
Taking $C=1$ and the standard value of $\gamma$, we get $w_+^{(B)}\approx 0.900$, that is greater than its analog $w_+^{(A)}$ in prescription A. 

Notice that the interval \eqref{inwbb} contains the relevant case where the second derivative of the potential is very close to zero and, hence, a region where ${}^{(B)}M_B^S$ is for sure negative does exist in this situation. Furthermore, this region reaches its largest possible extension when $\bar{W}^B_{,\tilde{\varphi}\tilde{\varphi}}= 0$, case in which Eq. \eqref{interx} becomes
\begin{align}\label{negB}
\dfrac{3}{4\pi G \sigma^2\rho_B}\bar{W}_B\in \left(\dfrac{1+\sqrt{2}}{3},1\right],
\end{align}
that approximately represents a $19.5\%$ of the space of the physical scalar field potentials.

On the other hand, the lower bound ${}^{(B)}M_B^S\geq Q_-$ ensures that the scalar effective mass at the bounce is positive when $Q_-$ is positive. As in prescription A, the values of $\bar{W}_B$ for which this occurs depend on the value of the argument of the square root in Eq. \eqref{xQ} corresponding to $Q_-$, that we call $T_-$, in comparison with a number of transitional values. It is convenient to rewrite the equality $T_-=D$ as a quadratic equation in the variable $w\in[0,\infty^+)$, with roots $w_{\pm}(D)$. It can be shown that there are no real roots for $D<\{2-9C^2(1+2\gamma^2)/[4(1+\gamma^2)]\}$, that is roughly $-0.370$ for $C=1$ and the standard value of $\gamma$. Therefore, if $C$ and $\gamma$ are close to their standard values, the equation $T_-=D$ has two real roots for any $D\geq 0$. We distinguish among four different situations: (i) if $w\in(w_-(0),w_+(0))$, both roots $x_\pm(Q_-)$ are complex and no new information is brought to light; (ii) if $w\in(w_-(1),w_-(0)]\cup [w_+(0),w_+(1))$, ${}^{(B)}M_B^S$ is nonnegative when $\bar{W}_B\in [x_-(Q_-),x_+(Q_-)]$; (iii) if $w\in[0,w_-(1)]\cup [w_+(1),w_+(4))$, ${}^{(B)}M_B^S$ is nonnegative when $\bar{W}_B\in [0,x_+(Q_-)]$; and (iv) if $w\geq w_+(4)$, ${}^{(B)}M_B^S$ is nonnegative for all $\bar{W}_B\in[0,4\pi G \sigma^2\rho_B/3]$. 

As commented in the previous subsection, in the case where the field potential is given by a mass term, $w$ is uniquely fixed by the mass of the scalar field. This mass (and therefore $w$) turns out to be extremely small in the cases of interest in LQC as regards the CMB \cite{CMBhybrid}. As a result, the physically interesting scenarios belong to case (iii), a fact that guarantees that the scalar effective mass at the bounce is positive in the sector of kinetic dominance. Indeed, when $\bar{W}^B_{,\tilde{\varphi}\tilde{\varphi}}$ is negligibly small, the effective mass at the bounce is positive for any $0\leq \bar{W}_B<4(1+\sqrt{2})\pi G \sigma^2\rho_B/9$.

To conclude this subsection, we compare the physical predictions of prescription A and B. In particular, it would be enlightening to see which region where $M_B^S$ is known to be negative is larger. In the limiting case of a vanishing $\bar{W}^B_{,\tilde{\varphi}\tilde{\varphi}}$, we have already verified that prescription B leads to a larger interval of field potentials for which the scalar effective mass at the bounce is necessarily negative [see Eqs. \eqref{negA} and \eqref{negB}]. However, this comparison can be extended to a more general scenario. In fact, one can wonder if $x_+(P_+)$ is greater than $x_+(Q_+)$ for any relevant second derivative of the potential [i.e., such that both $x_+(P_+)$ and $x_+(Q_+)$ are smaller that $4\pi G \sigma^2\rho_B/3$] and not only for a vanishing one. If the answer is in the affirmative, then $x_+(P_+)-x_+(Q_+)$ must be greater than zero, requirement which is equivalent to
\begin{align}\label{ineqw}
1+\sqrt{33+3w^2+12\sqrt{6}C\sqrt{\dfrac{1+2\gamma^2}{1+\gamma^2}}w}-\sqrt{32+6w^2+12\sqrt{6}C\sqrt{\dfrac{1+2\gamma^2}{1+\gamma^2}}w}>0.
\end{align}
This inequality is satisfied when $w^2=0$. Additionally, it can be shown that the left-hand side decreases as $w$ increases, reaching zero for a certain value $\bar{w}$ of $w$. 
If $\bar{w}$ were greater than the smallest of $w_+^{(A)}$ and $w_+^{(B)}$, then we would conclude that $x_+(P_+)>x_+(Q_+)$ for all relevant values of the second derivative of the field potential [actually, we know that $w_+^{(A)}<w_+^{(B)}$ for $C=1$ and the standard value of the Immirzi parameter, and hence the same will happen for values of $C$ and $\gamma$ that do not differ much from those]. From our above inequality \eqref{ineqw}, we realize that $\bar{w}$ verifies $-9\bar{w}^4+24\bar{w}^2+48\sqrt{6}C\sqrt{(1+2\gamma^2)/(1+\gamma^2)}\bar{w}+128=0$. This means that $\bar{w}$ is greater than one: indeed, the first and last terms already require that $\bar{w}>1$ and the remaining terms only increase the value of $\bar{w}$. Therefore, one can check that $\bar{w}$ is certainly greater than $w_+^{(A)}$ and $w_+^{(B)}$. As a result, we conclude that $x_+(P_+)>x_+(Q_+)$, not only in the limit of vanishing $\bar{W}^B_{,\tilde{\varphi}\tilde{\varphi}}$, but also for any relevant finite value.

In summary, although the scalar and tensor effective masses at the bounce are positive in the physically interesting cases (where $\bar{W}_B$ and $\bar{W}^B_{,\tilde{\varphi}\tilde{\varphi}}$ are small), prescription B leads to a scalar effective mass at the bounce that we can assure is negative in a region larger than its analog region for prescription A. Hence, prescription A is found to lead to effective masses at the bounce which are more appealing, in the sense that they are positive in a less restricted sector of scalar field potentials at the bounce.

\subsection{Dressed metric formalism}

We will evaluate now at the bounce the tensor and scalar effective masses in the dressed metric approach and discuss their positivity, in an analysis similar to that of the previous subsections. For this reason, we will focus mainly on the differences that arise in the process and provide the results for comparison with the ones of the hybrid approach.

On the one hand, the tensor effective mass at the bounce can be obtained from Eq. \eqref{MTD} by using Eqs. \eqref{dS1}-\eqref{dS4}:
\begin{align}
{}^{(D)}M_B^T=-4\pi G\dfrac{1+2\gamma^2}{1+\gamma^2}V^{2/3}_B(\rho_B+P_B)=-\dfrac{16\pi^2G^2}{3}\dfrac{1+2\gamma^2}{1+\gamma^2}\dfrac{(\pi_{\tilde{\varphi}}^B)^2}{V_B^{4/3}}.
\end{align}
In the last equality, we have employed the definitions \eqref{rho} and \eqref{P} of the matter energy density and the pressure in terms of their kinetic and potential contributions. It is manifest that the above expression is negative. This is in sharp contrast with the result obtained within the hybrid approach, where the tensor effective mass at the bounce \eqref{MThybridB} is positive in kinetically dominated regimes. This negativity of the tensor effective mass was also found in Ref. \cite{positividad} for the standard regularization of LQC. The result is hence robust against the ambiguity that affects the regularization adopted in the homogeneous Hamiltonian constraint.

On the other hand, using Eq. \eqref{UD}, the scalar effective mass at the bounce can be written as
\begin{align}
{}^{(D)}M_B^S=\dfrac{8\pi G}{3}V_B^{2/3}&\left\{-3\dfrac{1+2\gamma^2}{1+\gamma^2}\rho_B+\dfrac{3}{8\pi G \sigma^2}\left[\bar{W}^B_{,\tilde{\varphi}\tilde{\varphi}}+\left(36+6\dfrac{1+2\gamma^2}{1+\gamma^2}\right)\bar{W}_B\right.\right. \nonumber\\
&-\left.\left.s4\sqrt{6\pi G}\dfrac{\bar{W}^B_{,\tilde{\varphi}}|\pi^B_{\tilde{\varphi}}|}{V_B\sqrt{\rho_B}}-36\dfrac{3}{4\pi G \sigma^2}\dfrac{\bar{W}_B^2}{\rho_B}\right]\right\}.
\end{align}
Once again, we restrict our study to scalar field potentials that satisfy conditions \eqref{cond1} and \eqref{cond2}. For these potentials, the scalar effective mass at the bounce is bounded above and below by two quadratic polynomials in $\bar{W}_B$:
\begin{align}
K_\pm=\dfrac{8\pi G}{3}V_B^{2/3}&\left\{-3\dfrac{1+2\gamma^2}{1+\gamma^2}\rho_B+\dfrac{3}{8\pi G \sigma^2}\left[\bar{W}^B_{,\tilde{\varphi}\tilde{\varphi}}+\left(36+6\dfrac{1+2\gamma^2}{1+\gamma^2}\right)\bar{W}_B\right.\right. \nonumber\\
&\pm\left.\left.\dfrac{3\sigma C}{\gamma\sqrt{1+\gamma^2} \sqrt{\Delta}}\sqrt{\bar{W}^B_{,\tilde{\varphi}\tilde{\varphi}}}-36\dfrac{3}{4\pi G \sigma^2}\dfrac{\bar{W}_B^2}{\rho_B}\right]\right\}.
\end{align}
Their roots, denoted by $x_\pm(K_+)$ and $x_\pm(K_-)$, are
\begin{align}\label{xK}
\dfrac{3}{4\pi G \sigma^2}x_\pm(K_\pm)=\dfrac{6+\dfrac{1+2\gamma^2}{1+\gamma^2}\pm\sqrt{\left(6-\dfrac{1+2\gamma^2}{1+\gamma^2}\right)^2+\dfrac{3}{\pi G \sigma^2}\dfrac{\bar{W}^B_{,\tilde{\varphi}\tilde{\varphi}}}{\rho_B}\pm\dfrac{9C}{\pi G \sigma \gamma \sqrt{1+\gamma^2}\sqrt{\Delta}}\dfrac{\sqrt{\bar{W}^B_{,\tilde{\varphi}\tilde{\varphi}}}}{\rho_B}}}{12}\rho_B,
\end{align}
with the plus sign inside the square root corresponding to $K_+$. Since $\bar{W}^B_{,\tilde{\varphi}\tilde{\varphi}}>0$ with our restrictions, the roots $x_\pm(K_+)$ are real. Then, the upper bound on ${}^{(D)}M_B^S$ implies a negative scalar effective mass at the bounce when
\begin{align}
\bar{W}_B\in\left[0,\dfrac{4\pi G \sigma^2}{3}\rho_B\right]\setminus\left(\left[x_-(K_+),x_+(K_+)\right]\cap\left[0,\dfrac{4\pi G \sigma^2}{3}\rho_B\right]\right).
\end{align}
If the second derivative of the scalar field potential at the bounce is small enough, as in the most interesting cases for the CMB in LQC, it is straightforward to see that both roots are positive and $x_-(K_+)<4\pi G \sigma^2\rho_B/3<x_+(K_+)$. Therefore, when $\bar{W}^B_{,\tilde{\varphi}\tilde{\varphi}}$ is close to zero, the mass is negative for all $\bar{W}_B\in[0,x_-(K_+))$. This interval is large when the second derivative of the scalar field potential is negligibly small, and includes then the regimes where the matter energy density is dominated by its kinetic contribution.

The only way of avoiding a negative effective mass when the potential at the bounce is negligible compared with $4\pi G \sigma^2\rho_B/3$ is to have a large enough $\bar{W}^B_{,\tilde{\varphi}\tilde{\varphi}}$, so that $x_-(K_+)<0$. From Eq. \eqref{xK}, it is easy to show that this condition is met if and only if
\begin{align}
\bar{W}^B_{,\tilde{\varphi}\tilde{\varphi}}>\pi G \sigma^2 \rho_B\left(\sqrt{24C^2+8\dfrac{1+2\gamma^2}{1+\gamma^2}}-2\sqrt{6}C\right)^2.
\end{align}
Taking $C=1$ and the standard value of the Immirzi parameter, we obtain that the second derivative of the field potential at the bounce must be greater than approximately $0.633 \pi G \sigma^2 \rho_B$ for the scalar effective mass not to be negative at the bounce in scenarios of full kinetic dominance.

If the roots of $K_-$ are real as well (something that certainly does happen for sufficiently small values of $\bar{W}^B_{,\tilde{\varphi}\tilde{\varphi}}$), the lower bound on ${}^{(D)}M_B^S$ implies that the scalar effective mass  is positive at the bounce when $K_-$ is positive. As in the hybrid approach, the interval of values of the scalar field potential for which we can ensure that ${}^{(D)}M_B^S$ is positive depends on the value of the argument of the square root in Eq. \eqref{xK} corresponding to $K_-$, that we denote by $J_-$. We write $J_--D$ as a quadratic polynomial in the variable $w\in[0,\infty^+)$, with real roots $w_{\pm}(D)$. This assumption about the reality of the roots is justified, as in previous cases, by focusing on values of $C$ and $\gamma$ that are close to the standard ones. The fact that $w_-(D)$ happens to be always negative and $w_+(D)$ only lies on the positive semiaxis for $D$ above a certain threshold implies that, unlike in the hybrid approach, only two possible situations can occur. Indeed, the analogs of cases (i) and (ii) [see Subsecs. \ref{sec:bounceA} and \ref{sec:bounceB}] are found to be impossible.  The remaining ones are:
(iii) if $w\in[0,w_+(D_+))$, where $D_+=[6+(1+2\gamma^2)/(1+\gamma^2)]^2$, then ${}^{(D)}M_B^S$ is nonnegative for $\bar{W}_B\in[x_-(K_-),4\pi G \sigma^2\rho_B/3]$; and (iv) if $w\geq w_+(D_+)$, ${}^{(D)}M_B^S$ is nonnegative for all $\bar{W}_B\in[0,4\pi G \sigma^2\rho_B/3]$.

As we have already discussed,  in the case where the scalar field potential is simply a mass term, the second derivative of the potential is constant and $w$ is fixed by the mass of the scalar field. The physical scenarios of interest are characterized by an extremely small value of $w$ \cite{CMBdressed} and, therefore, they belong to case (iii): the mass is ensured to be positive only when the matter energy density at the bounce is dominated by its potential contribution. Consequently, we realize that $M_B^S$ can never be positive in the dressed metric formalism when \emph{both} $\bar{W}_B$ and $\bar{W}^B_{,\tilde{\varphi}\tilde{\varphi}}$ are (negligibly) small.

\section{Effective masses in the asymptotic de Sitter regime}\label{sec:dS}

In this section we will analyze the scalar and tensor effective masses corresponding to the hybrid and dressed metric approaches to LQC in the asymptotic de Sitter region of the prebounce branch. Once the masses are appropriately evaluated asymptotically, we will perform a separate study of their behavior and positivity, so as to compare the physical predictions corresponding to each approach in this regime that emerges as a result of the Dapor--Liegener regularization.

According to expression \eqref{rho}, there exist two situations in which a vanishing matter energy density is reached. One of them corresponds to a vanishing Hubble parameter, situation that occurs at large volumes in the FLRW branch, whereas the other involves a limit with constant Planckian Hubble parameter, that is clearly the case we are interested in. This limit is defined by $b\to b_0>0$ with
\begin{align}
\sin^2b_0=\dfrac{1}{1+\gamma^2}.
\end{align}
Thus, recalling our comment preceding Eq. \eqref{bB},
\begin{align}
b_0=\sin^{-1}\left(\dfrac{1}{\sqrt{1+\gamma^2}}\right).
\end{align}

On the one hand, a power expansion around $b=b_0$ confirms that the matter energy density does vanish asymptotically,
\begin{align}\label{rhodS}
\rho=\dfrac{3}{8\pi G \gamma^2(1+\gamma^2)\Delta}\left[-2\gamma+(1-5\gamma^2)(b-b_0)\right](b-b_0)+\mathcal{O}\left[(b-b_0)^3\right],
\end{align}
where the symbol $\mathcal{O}[\cdot]$ stands for terms of the order of the argument. On the other hand, taking the limit $b\to b_0$ in the square of the Hubble parameter [essentially given by the square of Eq. \eqref{V'}], it is immediate to verify that
\begin{align}
\lim_{b\to b_0}\left(\dfrac{V'}{3V^{4/3}}\right)^2=\dfrac{1}{4\gamma^2\Delta}\sin^2 2b_0\left[1-2(1+\gamma^2)\sin^2b_0\right]^2=\dfrac{\Lambda}{3},
\end{align}
where we have defined the  emergent cosmological constant of Planck order
\begin{align}
\Lambda=\dfrac{3}{(1+\gamma^2)^2\Delta}.
\end{align}

\subsection{Prescription A}\label{sec:dSA}

In order to be able to characterize the properties of the effective masses by analytical means in the de Sitter regime, we are going to restrict our attention to a class of scalar field potentials with a certain behavior in the asymptotic region, similarly to what we did in the discussion at the bounce. In the first place, we will focus our analysis on potentials that are asymptotically nonnegative. In that case, it is immediate to see that, in the asymptotic limit, both the potential and kinetic contributions to the matter energy density must vanish. Indeed, since $\rho$ is zero in this limit, the fact that the sum of the two nonnegative quantities $\pi_{\tilde{\varphi}}^2/(2V^2)$ and $9\bar{W}/(16\pi^2G^2\sigma^2)$ vanishes implies that both quantities must be zero. As a particular consequence, the scalar field must go to a zero of the potential, $\tilde{\varphi}_0$.

Additionally, we consider potentials with a first derivative that, in the asymptotic region, becomes much smaller than $\sqrt{\Lambda}\pi_{\tilde{\varphi}}/V$. Then, we can use the dynamical equations of effective LQC for the homogeneous background (recall that, at our order of perturbative truncation, the backreaction can be ignored in the calculation of the masses) in order to prove that, in the asymptotic regime, $\pi_{\tilde{\varphi}}/V$ and the scalar field grow exponentially in the proper time $t$. In more detail, once the backreaction is neglected, the effective equations of motion read
\begin{align}
\dfrac{dV}{dt}&=-\dfrac{3\gamma}{2\sqrt{\Delta}}V\sin 2b \left(1-\dfrac{1+\gamma^2}{\gamma^2}\cos 2b\right),\label{dVdt}\\
\dfrac{db}{dt}&=-3\gamma\sqrt{\Delta}\left(\dfrac{4\pi G }{3}\right)^2\left(\dfrac{\pi_{\tilde{\varphi}}}{V}\right)^2,\label{dbdt}\\
\dfrac{d\tilde{\varphi}}{dt}&=\dfrac{4\pi G }{3}\dfrac{\pi_{\tilde{\varphi}}}{V},\label{dphidt}\\
\dfrac{d\pi_{\tilde{\varphi}}}{dt}&=-\dfrac{3}{4\pi G \sigma^2}V\bar{W}_{,\tilde{\varphi}}.\label{dpidt}
\end{align}
From these, it is straightforward  to see that
\begin{align}
\dfrac{d}{dt}\left(\dfrac{\pi_{\tilde{\varphi}}}{V}\right)=-\dfrac{3}{4\pi G \sigma^2}\bar{W}_{,\tilde{\varphi}}+\dfrac{3\gamma}{2\sqrt{\Delta}}\sin 2b\left(1-\dfrac{1+\gamma^2}{\gamma^2}\cos 2b\right)\dfrac{\pi_{\tilde{\varphi}}}{V},
\end{align}
that asymptotically goes to
\begin{align}
\bigg[\dfrac{d}{dt}\left(\dfrac{\pi_{\tilde{\varphi}}}{V}\right)\bigg]_{0}=-\dfrac{3}{4\pi G \sigma^2}(\bar{W}_{,\tilde{\varphi}})_0+\sqrt{3\Lambda}\left(\dfrac{\pi_{\tilde{\varphi}}}{V}\right)_0,
\end{align}
where the subindex $0$ denotes the limit in the asymptotic de Sitter regime. Therefore{, if} according to our restrictions we have
\begin{align}\label{conddS}
\dfrac{3}{4\pi G \sigma^2}\left|(\bar{W}_{,\tilde{\varphi}})_0\right|\ll\sqrt{3\Lambda}\left|\left(\dfrac{\pi_{\tilde{\varphi}}}{V}\right)_0\right|,
\end{align}
then the momentum of the scalar field over the physical volume grows exponentially with the proper time in the de Sitter regime, as we have stated:
\begin{align}\label{asymppiV}
\dfrac{\pi_{\tilde{\varphi}}}{V}\approx C_{dS}e^{\sqrt{3\Lambda}t},
\end{align}
where $C_{dS}$ is an integration constant. Notice that, since this is precisely the behavior of $1/V$ in this region [see Eq. \eqref{dVdt}], the momentum of the zero mode of the scalar field has to be asymptotically constant. Furthermore, given that the derivative of the scalar field with respect to the proper time is equal to $\pi_{\tilde{\varphi}}/V$ up to a positive multiplicative constant, we conclude that the field grows exponentially as well in the asymptotic past. In fact, in this asymptotic regime,
\begin{align}\label{asympphi}
\tilde{\varphi}-\tilde{\varphi}_0\approx \dfrac{4\pi G }{3\sqrt{3\Lambda}}C_{dS}e^{\sqrt{3\Lambda}t}.
\end{align}
As a result, the ratio of the field and $\pi_{\tilde{\varphi}}/V$ is asymptotically constant:
\begin{align}
\lim_{b\to b_0}\dfrac{(\tilde{\varphi}-\tilde{\varphi}_0)V}{\pi_{\tilde{\varphi}}}=\dfrac{4\pi G }{3\sqrt{3\Lambda}}.
\end{align}

The validity of this result is subject to the verification of the approximation \eqref{conddS}. Such an approximation can be justified e.g. in the case that the scalar field potential is simply given by a mass term, namely $\bar{W}=\sigma^2 m^2(\tilde{\varphi}-\tilde{\varphi}_0)^2/2$, where $m$ is the mass of the scalar field. For this potential, Eq. \eqref{conddS} adopts the form 
\begin{align}
\lim_{b\to b_0}\dfrac{3m^2}{4\pi G}\left|\dfrac{(\tilde{\varphi}-\tilde{\varphi}_0)V}{\pi_{\tilde{\varphi}}}\right|\ll\sqrt{3\Lambda} \Leftrightarrow m^2\ll 3\Lambda.
\end{align}
Hence, we realize that the condition that the first derivative of the potential be much smaller than $\sqrt{\Lambda}\pi_{\tilde{\varphi}}/V$ in the de Sitter regime is valid for the case of a mass term in the phenomenologically favored scenarios, where the mass $m$ is very small \cite{CMBhybrid,CMBdressed} (in particular, compared with the square root of the Planckian cosmological constant).

Moreover, since in the studied asymptotic de Sitter region the scalar field behaves like the square root of the kinetic contribution to the matter energy density ($\pi_{\tilde{\varphi}}/V$, up to a factor), with a proportionality constant that is of the order of the inverse of the square root of the emergent cosmological constant, it is now immediate to see that the scalar field potential turns out to be negligible with respect to the matter energy density provided that it varies with the scalar field faster than its square, or that it varies with the square but multiplied by  a factor that is much smaller than the cosmological constant. The latter condition is met in the case of the mass term under consideration, where the field potential is quadratic in the scalar field but the square of the mass is much smaller than $\Lambda$. Thus, in the following discussion, we neglect the asymptotic contribution of the field potential to the matter energy density.

It is also worth noticing that, if the upper bound on $|\bar{W}_{,\tilde{\varphi}}|$ that we employed at the bounce \eqref{cond2} is satisfied asymptotically as well, the restriction that this derivative be much smaller than the square root of the kinetic energy turns out to hold, provided that (a) the scalar field potential itself be much smaller than the kinetic energy, something that can be granted for the family of potentials described in the above paragraph, and (b) the second derivative of the potential be asymptotically finite, restriction that seems very reasonable to impose. Then, for potentials in the mentioned family and for which $\bar{W}_{,\tilde{\varphi}\tilde{\varphi}}$ is finite in the de Sitter limit,  given our dynamical equations it is enough to impose just one of the two conditions: the analog of the upper bound \eqref{cond2} on $|\bar{W}_{,\tilde{\varphi}}|$ in the asymptotic region, or the alternative upper bound \eqref{conddS}.

To support the conclusions of our analysis, we have integrated numerically the effective dynamical equations for the background in the case of a scalar field of mass $m=1.2\cdot 10^{-6}$ in geometrical natural units (i.e., with $G=1$), that is a typical value leading to power spectra of interest for the CMB in LQC \cite{Universe}. In order to perform the numerical integration, we first need to provide a set of initial conditions. Usually, these are given at the bounce, that yields a privileged spatial section (of minimum volume) and a natural choice of initial time. In particular, employing again natural units, we have taken the initial condition $\tilde{\varphi}_B=0.97$ at the bounce \cite{Universe}, setting the global volume scale of the flat FLRW cosmology so that $V_B=(2\pi)^3$. In addition, we recall that the variable $b$ adopts the value $b_B=\sin^{-1}[{1}/{\sqrt{2(1+\gamma^2)}}]$ at the bounce, and that the momentum of the scale factor is then determined by the vanishing of the effective homogeneous Hamiltonian constraint, that gives
$\pi_{\tilde{\varphi}}^B=V_B [24\pi \rho_B- 9m^2 (\tilde{\varphi}_B-\tilde{\varphi}_0)^2]^{1/2}/(4\pi )$.

\begin{figure}[h!]
\centering
\includegraphics[scale=1]{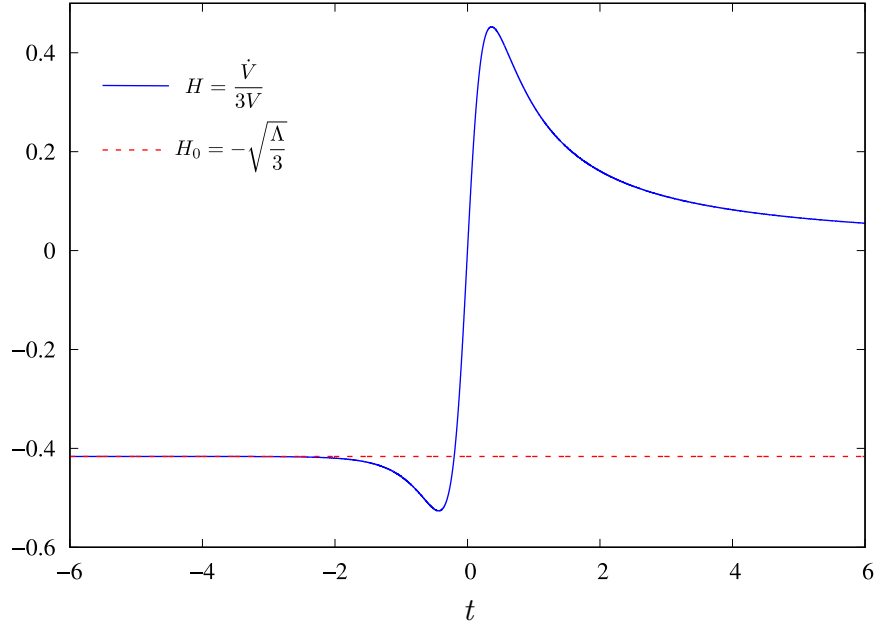}
\caption{Hubble parameter as a function of the proper time.}
\label{fig:hubble}
\end{figure}

\begin{figure}[h!]
\centering
\includegraphics[scale=1]{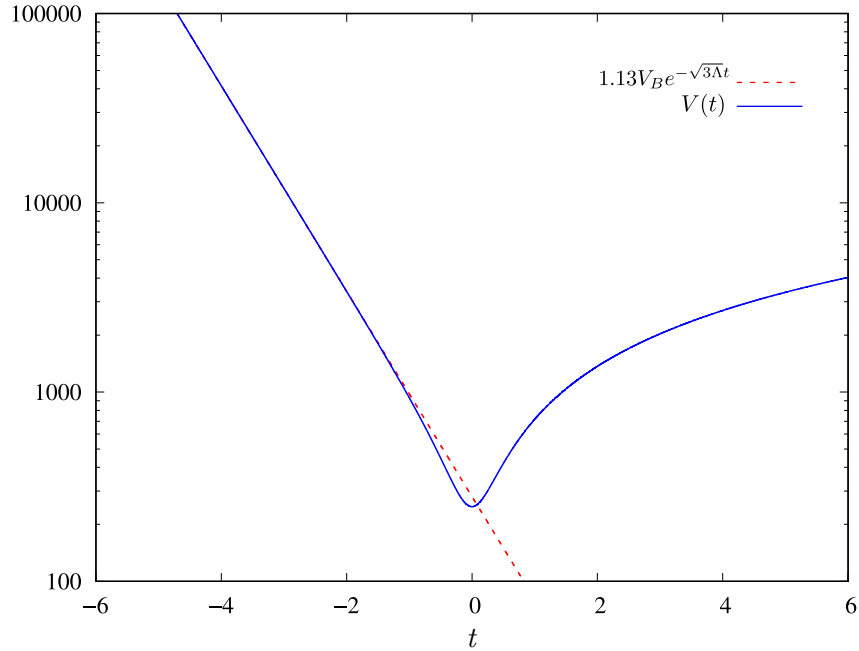}
\caption{Volume in logarithmic scale as a function of the proper time.}
\label{fig:V}
\end{figure}

\begin{figure}[h!]
\centering
\includegraphics[scale=1]{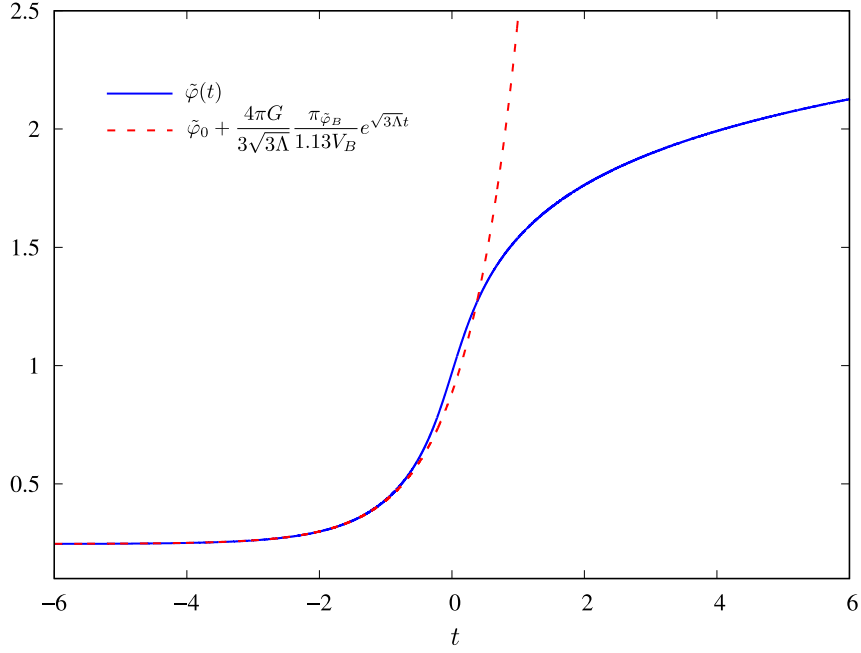}
\caption{Scalar field as a function of the proper time.}
\label{fig:phi}
\end{figure}

\begin{figure}[h!]
\centering
\includegraphics[scale=1]{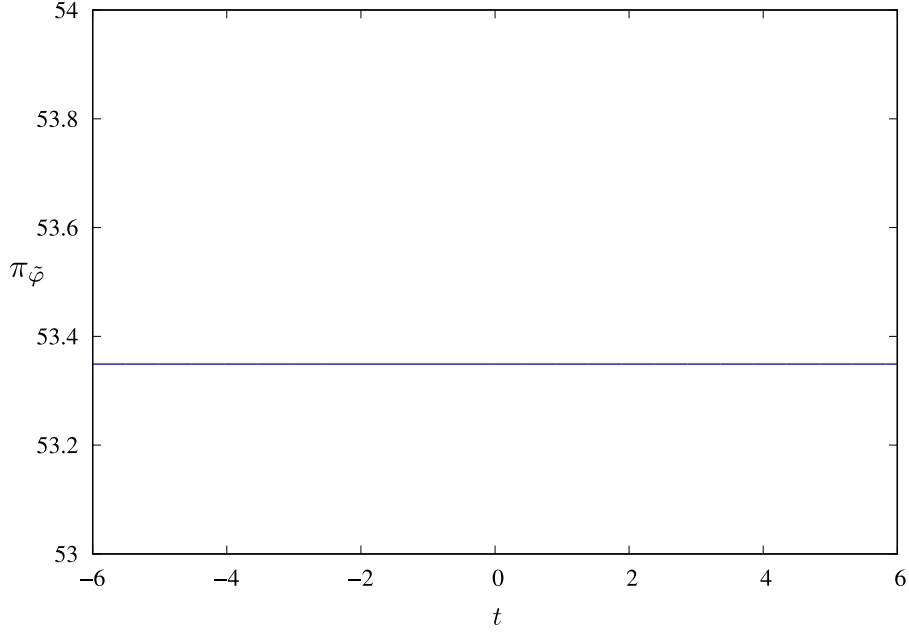}
\caption{Momentum of the scalar field as a function of the proper time.}
\label{fig:piphi}
\end{figure}

\begin{figure}[h!]
\centering
\includegraphics[scale=1]{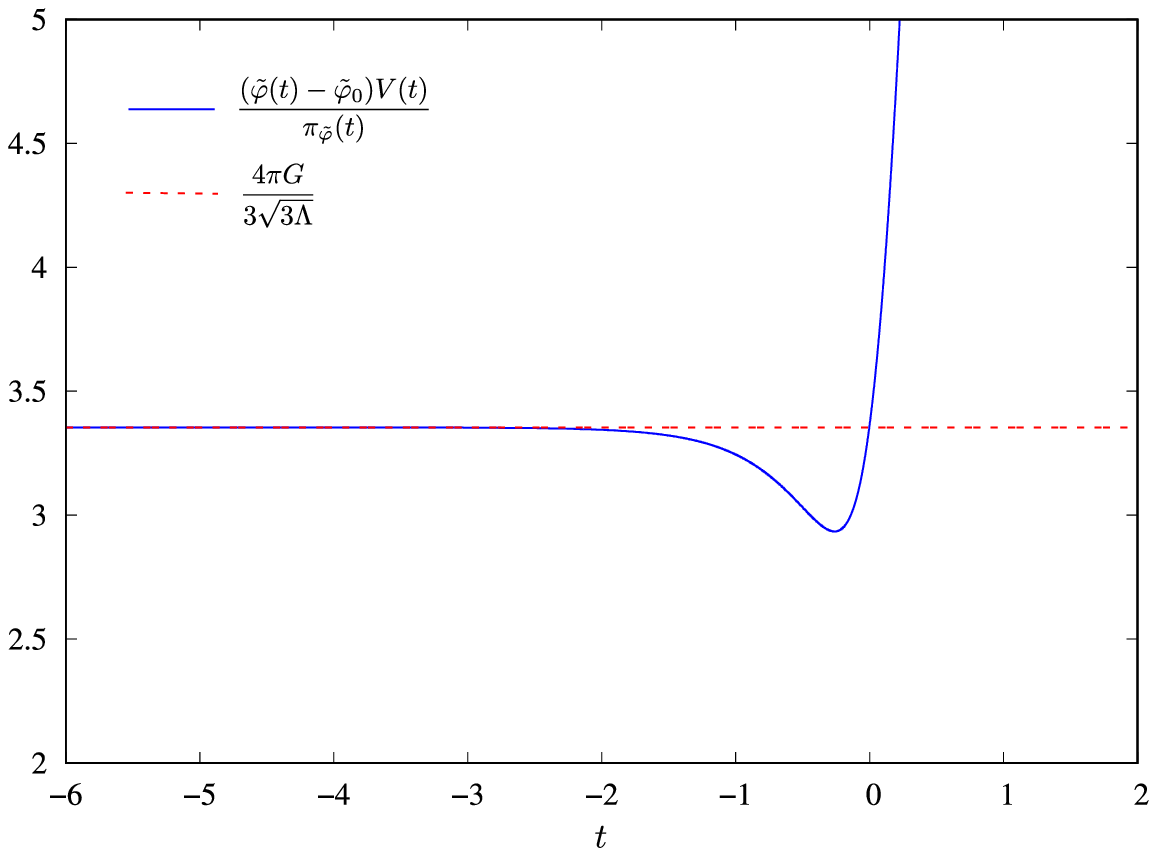}
\caption{Ratio $(\tilde{\varphi}-\tilde{\varphi}_0)V/\pi_{\tilde{\varphi}}$ as a function of the proper time.}
\label{fig:quot}
\end{figure}

The result of the numerical integration of Eqs. \eqref{dVdt}-\eqref{dpidt} with these initial conditions is shown  in Figs. \ref{fig:hubble}-\ref{fig:quot}. In the numerics, performed in natural units, we have set the coordinate length $l_0$ of the fundamental cycles of the spatial sections equal to $2\pi$. The bounce has been located at vanishing proper time $t$. Fig. \ref{fig:hubble} shows that the prebounce branch approaches a de Sitter phase very rapidly. In fact, the Hubble parameter attains its predicted asymptotic value only a few Planck times away from the bounce, a fact which implies that we can safely study the asymptotic de Sitter region by considering the behavior of the dynamical variables all along the evolution from the far past to a few Planck times before the big bounce. Figs. \ref{fig:V} and \ref{fig:phi} confirm that the physical volume and the zero mode of the scalar field behave in the asymptotic past as expected: the field grows exponentially forward in the proper time, while the physical volume contracts at the same rate. This leads to a momentum of the scalar field that is approximately constant, not only asymptotically, but also across the bounce, as revealed by Fig. \ref{fig:piphi}. This can be understood by realizing that the momentum of the scalar field is a constant of motion in the case of a massless scalar field. Hence, the smallness of the phenomenologically preferred mass only breaks this symmetry slightly, producing a very slow variation of $\pi_{\tilde{\varphi}}$ in the case under consideration. Finally, Fig. \ref{fig:quot} corroborates that, in the scenarios of phenomenological interest, the ratio $(\tilde{\varphi}-\tilde{\varphi}_0)V/\pi_{\tilde{\varphi}}$ goes asymptotically to the predicted constant value and, therefore, ignoring the first derivative of the field potential with respect to $\sqrt{\Delta}\pi_{\tilde{\varphi}}/V$ is a good approximation.

On the light of the conclusions of our analytical and numerical study, in the following we will asymptotically ignore the scalar field potential in comparison to the kinetic energy and, thus, to the matter energy density itself. The same holds true as regards the first derivative of the potential with respect to the square root of the matter energy density. Then, we obtain that the tensor effective mass in the de Sitter region is [see Eqs. \eqref{exprMT} and \eqref{rhodS}]
\begin{align}\label{MTdShybrid}
M_{dS}^T=\dfrac{8\pi G}{3}V^{2/3}\rho+\mathcal{O}\left[\bar{W}\right]=\dfrac{V^{2/3}}{\gamma^2(1+\gamma^2)\Delta}\left[-2\gamma+(1-5\gamma^2)(b-b_0)\right](b-b_0)+\mathcal{O}\left[\bar{W},(b-b_0)^3\right].
\end{align}
Notice that this mass is proportional to $V^{2/3}$. This is exactly the same power of the volume that appears in the Mukhanov--Sasaki potential \eqref{UA}. So, the relative smallness or dominance of that potential in the asymptotic region is independent of this factor. In addition, the asymptotic growth of $V^{2/3}$ backwards in time cannot compensate the vanishing of the matter energy density in the tensor effective mass, because the energy density can be approximated by its kinetic part that vanishes asymptotically as the square inverse of the volume, since the momentum of the scale factor remains approximately constant, as we have seen.

The fact that the tensor effective mass in this approximation is given by the matter energy density multiplied by a strictly positive factor guarantees the nonnegativity of the mass. Equivalently, we see that the dominant linear term in $b-b_0$ of the tensor effective mass is always positive, since $\gamma>0$ and we recall that $b-b_0\leq 0$ (by virtue of the monotonically decreasing nature of the connection-like variable in the solution under consideration). Additionally, for the standard value of the Immirzi parameter, we find that the quadratic term does not destroy the positivity of the linear one either, not only for small values of $b-b_0$, but in general (at least, as long as the scalar field potential can still be neglected at this level). Indeed, the quadratic term is positive provided that $\gamma^2<1/5$, inequality that $\gamma_{\rm stand}^2\approx 0.0564$ clearly satisfies.

Let us analyze now the Mukhanov--Sasaki potential \eqref{UA} in the asymptotic region. We recall that it receives contributions from the field potential, its square, its first derivative, and its second derivative. From our previous considerations, it is clear that the contributions of the field potential and its square vanish in the asymptotic past. Moreover, they are asymptotically much smaller than any term of the order of the matter energy density. As a result, only the contributions of the derivatives of the field potential remain. If we allow for an asymptotically nonvanishing second derivative of the field potential, the first term in Eq. \eqref{UA} will certainly contribute to the scalar effective mass in the de Sitter regime. As for the contribution of $\bar{W}_{,\tilde{\varphi}}$, two different situations must be told apart within the family of scalar field potentials that we are considering. If the potential varies asymptotically with the scalar field faster than its square, its first derivative should vary at least faster than the field itself, and therefore should vanish in the asymptotic limit faster than $\rho^{1/2}$. It is then immediate to realize that the third term in Eq. \eqref{UA} would vanish in the asymptotic region, so that one can neglect it compared to the contribution of the second derivative of the potential: ${}^{(A)}\mathcal{U}_{dS}=V^{2/3}(\bar{W}_{,\tilde{\varphi}\tilde{\varphi}})_0/\sigma^2$. On the other hand, if the potential varied asymptotically precisely with the square of the field (as in the case of a mass term), its first derivative would go as $\rho^{1/2}$, a behavior which leads to a third term in Eq. \eqref{UA} that no longer vanishes, but becomes constant asymptotically. Indeed, in the case of a mass term as the one considered above,
\begin{align}
\lim_{b\to b_0}\dfrac{\bar{W}_{,\tilde{\varphi}}\pi_{\tilde{\varphi}}}{V\rho}=\dfrac{2\sigma^2m^2}{\sqrt{3\Lambda}}=\dfrac{2}{3}\sigma^2m^2(1+\gamma^2)\sqrt{\Delta},
\end{align}
where we have used the definition of the matter energy density \eqref{rho}, ignored the scalar field potential in it, and employed Eqs. \eqref{asymppiV} and \eqref{asympphi}. Thus, in the case of a mass term, the contributions of $\bar{W}_{,\tilde{\varphi}}$ and $\bar{W}_{,\tilde{\varphi}\tilde{\varphi}}$ are of the same order and the Mukhanov--Sasaki potential \eqref{UA} is asymptotically given by
\begin{align}
{}^{(A)}\mathcal{U}_{dS}\approx m^2V^{2/3}\left[1+\dfrac{2(1+\gamma^2)}{\gamma}\sin 2b_0\right]=5m^2V^{2/3}.
\end{align}
The same behavior is found for any scalar field potential that varies quadratically with the field asymptotically, replacing $m^2$ with $(\bar{W}_{,\tilde{\varphi}\tilde{\varphi}})_0/\sigma^2$.

In conclusion, since the tensor effective mass vanishes asymptotically, the introduction of the Mukhanov--Sasaki potential entails the positivity of the scalar effective mass, as long as the second derivative of the scalar field potential is asymptotically greater than zero. We remark that, in the special case of potentials that are asymptotically quadratic in the scalar field, like e.g. a mass term, the contribution of the first derivative of the field potential must also be taken into consideration and strengthens the positivity of the scalar effective mass in the asymptotic de Sitter regime, given our assumption that $(\bar{W}_{,\tilde{\varphi}\tilde{\varphi}})_0>0$. 

\subsection{Prescription B}\label{sec:dSB}

Since the tensor effective mass does not depend on the prescription adopted for the representation of $1/\pi_{\tilde{\alpha}}^2$, our conclusions about its behavior and positivity presented in Subsec. \ref{sec:dSA} remain valid [see Eq. \eqref{MTdShybrid} and the discussion in the paragraph below it]. And, furthermore, if the restrictions on the scalar field potential introduced in the previous subsection are verified, all the terms in ${}^{(B)}\mathcal{U}$ that differ from ${}^{(A)}\mathcal{U}$ are easily seen to be negligible in the asymptotic region. In fact, all but the first term in Eq. \eqref{UB} can be proven much smaller than the matter energy density  in the de Sitter region, up to factors of the order of $V^{2/3}$. Hence, only the contribution of the second derivative of the scalar field potential survives. As a consequence of this fact, we can assure that the scalar effective mass in prescription B is also asymptotically strictly positive provided that $(\bar{W}_{,\tilde{\varphi}\tilde{\varphi}})_0>0$ (a condition that we know that holds in the case of a mass term).

\subsection{Dressed metric formalism}

It is also possible to express the tensor effective effective mass of the dressed metric approach, given by Eq. \eqref{MTD}, in terms of the energy density and the field potential in the de Sitter branch. This can be done by means of Eq. \eqref{sin2brho} and Eqs. \eqref{dS1}-\eqref{dS4}, identified with the subindex $+$. Since we are interested in the vicinity of the de Sitter regime, we can expand the result in a power series of the ratio $\rho/\rho_B$, which is very small when the de Sitter region is approached. The result of this computation is
\begin{align}
{}^{(D)}M_{dS}^T=-2V^{2/3}\left[\dfrac{\Lambda}{3}\left(1-\dfrac{1-5\gamma^2}{8\gamma^2}\dfrac{\rho}{\rho_B}\right)+\dfrac{3\bar{W}}{\sigma^2(1+\gamma^2)}\left(1-5\gamma^2-\dfrac{4-3\gamma^2}{2}\dfrac{\rho}{\rho_B}\right)\right]\left\{1+\mathcal{O}\left[\dfrac{\rho^2}{\rho_B^2}\right]\right\}.\label{mdseq}
\end{align}

We consider again the same restrictions on the class of scalar field potentials under study that were employed in the hybrid case for the analysis of the de Sitter region, namely: the potential is asymptotically nonnegative (so that both its contribution to the matter energy density and the kinetic one vanish in the de Sitter limit), the potential is asymptotically much smaller than the kinetic contribution (and, thus, than the energy density itself), and the first derivative of the potential is asymptotically much smaller than $\sqrt{\Lambda}\pi_{\tilde{\varphi}}/V$ (and, therefore, than the square root of the energy density). It is also important to bear in mind the considerations about when these restrictions are fulfilled, discussed in Subsec. \ref{sec:dSA}. In principle, we allow for the possibility of an asymptotically nonnegligible second derivative of the scalar field potential. In this situation, we may  ignore the terms proportional to $\bar{W}$ in Eq. \eqref{mdseq}, since those that are linear in $\rho$ dominate:
\begin{align}
{}^{(D)}M_{dS}^T=-\dfrac{2\Lambda}{3}V^{2/3}\left(1-\dfrac{1-5\gamma^2}{8\gamma^2}\dfrac{\rho}{\rho_B}\right)+\mathcal{O}\left[\bar{W},\dfrac{\rho^2}{\rho_B^2}\right].
\end{align}
We clearly see that, in the vicinity of de Sitter regime, where $\rho$ is much smaller than $\rho_B$, the tensor effective mass in the dressed metric formalism is negative. The first correction to the limit value, linear in $\rho/\rho_B$, is positive for $\gamma^2<1/5$, and thus in the case of the standard value of the Immirzi parameter. Hence, the negativity of the tensor effective mass diminishes away from the asymptotic de Sitter regime.

On the other hand, similar arguments to those commented in the previous subsections show that every term in the Mukhanov--Sasaki potential \eqref{UD} is asymptotically negligible compared to the first one. Then, we find the same type of asymptotic behavior that we found within the hybrid approach:
\begin{align}
{}^{(D)}\mathcal{U}_{dS}\approx V^{2/3}\dfrac{\left(\bar{W}_{,\tilde{\varphi}\tilde{\varphi}}\right)_0}{\sigma^2}.
\end{align}
As a result, the positivity may be restored at the level of the scalar effective mass thanks to the contribution of the Mukhanov--Sasaki potential, provided that $\left(\bar{W}_{,\tilde{\varphi}\tilde{\varphi}}\right)_0$ is positive and large enough. However, this cannot be achieved in the physically interesting scenarios where the scalar field is only subject to a mass term, with a mass much smaller than the square root of the emergent cosmological constant. Indeed, in that case $\bar{W}_{,\tilde{\varphi}\tilde{\varphi}}/\sigma^2$ is constant and equal to $m^2$, giving a negligible contribution compared with the limit value of the tensor effective mass and, thus, proving insufficient to attain a strictly positive scalar effective mass in the asymptotic de Sitter region.

\section{Discussion and conclusions}\label{sec:conclusion}

In the past years, a special effort has been devoted to the investigation of the mathematical ambiguities that affect the formalism of LQC, as a theory that attempts to apply the techniques of LQG in cosmology. More specifically, a particular attention has been paid to the definition of the Hamiltonian in cosmological spacetimes. Indeed, the publication of a work by Dapor and Liegener, where a new regularization procedure for the Hamiltonian was put forward \cite{DL}, sparked a considerable number of studies analyzing its consequences and comparing its physical predictions with those of the so-far standard regularization in LQC, in the search for robust results that may point at genuine features of the cosmological dynamics within the full theory of LQG. Although the resolution of the initial cosmological singularity was still present, qualitative differences with respect to the standard case were pointed out: instead of a symmetric bounce that joins two universes that behave classically at large volumes, one of them is now replaced by an asymptotically de Sitter branch with a curvature of the Planck order. Hence, even though some details of the bouncing mechanism may be regularization dependent, there are solid reasons to trust the singularity resolution itself.

In the same spirit, the new regularization procedure has been thoroughly studied in a variety of contexts, so as to discern how a different regularization of the geometry alters the traditional results in the field \cite{Paramc1,Paramc2,genericness,Agullo,Haro,MMODL}. Additionally, the Dapor--Liegener model has been implemented in more general scenarios \cite{DLBI}, in order to achieve a better understanding of how this proposal differs from the standard one in systems possessing less symmetries. From this point of view, the most interesting symmetry to dispose of is homogeneity, since inhomogeneities should play a fundamental role in the formation of the LSS and the anisotropies of the CMB. Therefore, it is from models that somehow include inhomogeneities that there is hope to extract falsifiable physical predictions that may put the theory to test. Consequently, an analysis of the effects of the selection of a different regularization scheme in inhomogeneous spacetimes seems to be in order.

In the framework of LQC two main paths have been followed for the inclusion of inhomogeneities in otherwise homogeneous cosmological models: the so-called hybrid \cite{inflationaryuniverse,inflationarymodel,MSvariables,GIper,quantumcorrectionsMSeq,Olm} and dressed metric \cite{dressed1,dressed2,dressed3,dressed4} approaches. Both proposals introduce the inhomogeneities perturbatively on a homogeneous and isotropic background and select different representations for the homogeneous and inhomogeneous sectors, arguing that there must exist a physical regime where the main quantum geometry effects are those that affect directly the homogeneous part of the system, whereas the perturbations can be treated by means of standard field theoretical techniques on the resulting background of quantum nature. Nonetheless, the two formalisms are constructed in different ways: while in the hybrid approach one regards the entire cosmological system as a whole and quantizes it accordingly (treating the homogeneous background and the perturbations on the same footing), the dressed metric formalism proposes a program consisting of two steps, dealing with the homogeneous sector first and then studying the propagation of the perturbations on a ``dressed'' background (with no backreaction of the perturbations). Hence, one should expect differences in the physical predictions of both formalisms when the quantum geometry effects are important. In this regard, the time-dependent masses that govern the dynamics of the perturbations seem especially appropriate to discuss potential discrepancies. In particular, the positivity of these time-dependent masses is important to pose well-defined initial conditions for the perturbations, since the oscillatory behavior of the perturbations for all wavelengths depends on this positive character at the end of the day. The analysis of this positivity is decidedly interesting in regimes where there exist physical motivations to set those initial conditions. A study and comparison of the time-dependent masses derived from the hybrid and dressed metric approaches, for backgrounds that follow the effective dynamics of LQC, were already performed at the bounce in Ref. \cite{positividad} employing the standard regularization scheme. The objective of this paper is to establish an analogous comparison when the Dapor--Liegener regularization is used to construct the Hamiltonian of the homogeneous cosmology. Besides, it seems appropriate to include in this comparison an analysis of the time-dependent masses, evaluated within effective LQC, in the vicinity of the asymptotic de Sitter regime. This positivity is important for the construction of adiabatic states for all wavelengths. If there exist obstructions to this construction, a naive characterization of the Bunch--Davies state as a state of infinite adiabatic order with the isometries of the de Sitter cosmology certainly will be compromised. Furthermore, one may wonder whether a non-Einsteinian behavior of the effective mass indicates that one should revisit the role of this state for asymptotic regimes out of general relativity like those emerging in LQC.

With these aims in mind, we have first introduced the main elements of the hybrid and dressed metric formalisms that are relevant to our discussion and written down the time-dependent masses that enter the equations of scalar and tensor perturbations, evaluating them on quantum background states that are peaked on trajectories of the effective dynamics of LQC. In the hybrid case, we have analyzed separately two prescriptions (called A and B), that are intimately related to the regularized quantum structure of the theory and that were introduced in Ref. \cite{DLhLQC}. We have evaluated the resulting effective masses in the regimes that seem more appealing, as far as the setting of initial conditions for the perturbations is concerned: the big bounce and the asymptotic de Sitter era.

In the hybrid case, we have shown that the tensor effective mass is strictly positive in a region that encompasses the scenarios of kinetic dominance at the bounce. In addition, we have seen that the Mukhanov--Sasaki potential at the bounce in this hybrid case exhibits a difference with respect to the one presented in Ref. \cite{positividad} which is worth mentioning: it contains a term that is proportional to the first derivative of the field potential that does not vanish at the bounce (in contrast with the situation found with the standard regularization of the homogeneous geometry). This complicates the subsequent analysis and requires the assumption of an upper bound on the absolute value of this term, in order to be able to proceed analytically. With this and other mild conditions on the scalar field potential, we have obtained some convenient upper and lower bounds on the scalar effective mass. Moreover, provided that the second derivative of the field potential is small enough, we have shown that there exists a region of the space of physical potentials where the scalar effective mass is ensured to be negative. This region, however, only contains scenarios where the matter energy density at the bounce is dominated by the potential contribution. Indeed, when the second derivative of the potential is sufficiently small, the scalar effective mass within the hybrid approach is also found to be positive on solutions with kinetic dominance at the bounce. This conclusion applies to both prescriptions A and B. The difference between them is that the application of prescription B results in a larger region of negative scalar effective masses, a fact which can be used to argue that prescription A is not only theoretically better motivated \cite{DLhLQC}, but it also leads to more appealing physical features. In the dressed metric case, on the other hand, we arrive at a strictly negative tensor effective mass at the bounce. This result is actually explained by the fact that the considered mass is the ratio of the second derivative of the scale factor and the scale factor itself, except for a sign. Thus, given that the scale factor as a function of time is concave at the bounce, it is unavoidable to get a negative value in effective LQC. Additionally, the scalar effective mass at the bounce is ensured to be negative in a considerably large sector of solutions that includes the kinetically dominated scenarios, at least for a small second derivative of the potential at the bounce. If this second derivative were large enough, however, the scalar effective mass could become positive at the bounce. This does not avoid the strict negativity of both masses in the physically interesting case where the scalar field potential and its second derivative are small. As a result, in the dressed metric approach one would encounter certain obstructions in the attempt to construct adiabatic states at the bounce by conventional procedures, something that generically can only be achieved for a restricted range of wavelengths if the effective masses are not positive.

Remarkably, the conclusions reached in this work about the positivity or negativity of the effective masses at the bounce in the sector of kinetic dominance coincide qualitatively in general terms  with the results of Ref. \cite{positividad} for the case of the standard regularization of the homogeneous Hamiltonian constraint in LQC, and in this sense we can say that these conclusions are robust against changes in the regularization scheme adopted for the Hamiltonian in LQC.

On the other hand, we have also carried out a similar analysis of the positivity of the masses in the asymptotic de Sitter branch. In order to evaluate the effective masses in this regime, we have introduced some reasonable restrictions on the scalar field potential that allow us to ignore its contributions and those of its first derivative as compared with the matter energy density and its square root, respectively. To motivate these conditions, we have discussed the asymptotic behavior of the effective equations of motion for the background and shown that, in particular, the mass potential satisfies all considered requirements. We have also integrated numerically the dynamical equations of the effective description of the homogeneous cosmology and verified that the introduced restrictions on the scalar field potential hold in the cases of interest, as regards the computation of the power spectrum of the CMB in LQC. In the hybrid case, prescriptions A and B lead to the same value of the tensor effective mass, that vanishes identically in the asymptotic past and is positive in its vicinity, given that it is proportional to the matter energy density. As far as the positivity of the asymptotic scalar effective mass is concerned, the asymptotic vanishing of the tensor effective mass entails that the limit behavior of the Mukhanov--Sasaki potential is of crucial importance. Our analysis shows that the positivity of the scalar effective mass is granted owing to the contribution of the Mukhanov--Sasaki potential if the second derivative is strictly positive in the de Sitter limit. On the other hand, for the dressed metric formalism we have shown that the tensor effective mass is negative asymptotically but becomes less negative away from the de Sitter regime. The contribution of the Mukhanov--Sasaki potential can make the scalar effective mass positive, but only provided that the second derivative of the scalar field potential is positive and large enough. This possibility is ruled out in the case of a massive scalar field for phenomenologically favored values of the field mass. 

On the light of these results, we conclude that the formalism of hybrid LQC leads to effective masses for the perturbations that overall display more attractive features, inasmuch as they are ensured to be positive both at the bounce and in the de Sitter limit, for the scenarios of physical interest and a large class of scalar field potentials with direct application, to which a mass term belongs. The effective masses derived from the dressed metric approach, in contrast, cannot be made positive in the phenomenologically favored scenarios, at least at the bounce and in the asymptotic de Sitter regime. As far as the tensor effective mass is concerned, it is strictly negative at the bounce and in the infinite past. The scalar effective mass, however, could become positive in presence of a large second derivative of the field potential, condition which clashes with the fact that the physically interesting field masses are typically very small. 

\acknowledgments
	
The authors are very grateful to B. Elizaga Navascu\'es for discussions. This work has been supported by Project. No. FIS2017-86497-C2-2-P of MICINN from Spain. A. Garc\'{\i}a-Quismondo acknowledges that the project that gave rise to these results received the support of a fellowship from ``la Caixa'' Foundation (ID 100010434). The fellowship code is LCF/BQ/DR19/11740028. G. S\'anchez P\'erez acknowledges support of the Grant No. CSIC JAEINT19\_EX\_0632.

\end{document}